\pdfoutput=1
\documentclass[sigconf]{acmart} 
\usepackage{verbatim}
\usepackage{amsmath,amssymb,amsfonts}
\usepackage{graphicx}
\usepackage{textcomp}
\usepackage{color}
\usepackage{subfigure} 
\usepackage{breakurl}
\usepackage{multirow}
\usepackage{diagbox}
\usepackage{makecell}
\usepackage{booktabs}
\usepackage[normalem]{ulem}
\usepackage{enumitem}

\newcommand{\mysize}{\fontsize{8}{8}\selectfont}

\newcommand{\etal}{{\textit{et al. }}}

\renewcommand\footnotetextcopyrightpermission[1]{} 
\AtBeginDocument{%
  \providecommand\BibTeX{{%
    \normalfont B\kern-0.5em{\scshape i\kern-0.25em b}\kern-0.8em\TeX}}}

\setcopyright{none}

\begin{document}
\fancyhead{}

\settopmatter{printacmref=false, printccs=true, printfolios=true} 






\title{Off-Path TCP Exploits of the Mixed IPID Assignment}


\author{Xuewei Feng$^{1}$, Chuanpu Fu$^{1}$, Qi Li$^{2,3}$, Kun Sun$^{4}$, and Ke Xu$^{1,3,5}$} 
\affiliation{
$^{1}$Department of Computer Science and Technology, Tsinghua University, Beijing, China\\
$^{2}$Institute for Network Sciences and Cyberspace, Tsinghua University, Beijing, China\\
$^{3}$Beijing National Research Center for Information Science and Technology (BNRist), Tsinghua  University, Beijing, China\\
$^{4}$Department of Information Sciences and Technology, CSIS, George Mason University\\
$^{5}$Peng Cheng Laboratory, China\\
\{fengxw18@mails, fcp20@mails, qli01@, xuke@\}tsinghua.edu.cn, ksun3@gmu.edu
}

\begin{abstract}
In this paper, we uncover a new off-path TCP hijacking attack that can be used to terminate victim TCP connections or inject forged data into victim TCP connections by manipulating the new mixed IPID assignment method, which is widely used in Linux kernel version 4.18 and beyond to help defend against TCP hijacking attacks. The attack has three steps. First, an off-path attacker can downgrade the IPID assignment for TCP packets from the more secure per-socket-based policy to the less secure hash-based policy, building a shared IPID counter that forms a side channel on the victim. Second, the attacker detects the presence of TCP connections by observing the shared IPID counter on the victim. Third, the attacker infers the sequence number and the acknowledgment number of the detected connection by observing the side channel of the shared IPID counter. Consequently, the attacker can completely hijack the connection, i.e., resetting the connection or poisoning the data stream.



We evaluate the impacts of this off-path TCP attack in the real world. Our case studies of SSH DoS, manipulating web traffic, and poisoning BGP routing tables show its threat on a wide range of applications.
Our experimental results show that our off-path TCP attack can be constructed within 215 seconds and the success rate is over 88\%. Finally, we analyze the root cause of the exploit and develop a new IPID assignment method to defeat this attack. We prototype our defense in Linux 4.18 and confirm its effectiveness through extensive evaluation over real applications on the Internet.
\end{abstract}




\maketitle

\textbf{\mysize ACM Reference Format:}\\
{\mysize Xuewei Feng, Chuanpu Fu, Qi Li, Kun Sun, and Ke Xu. 2020. Off-Path TCP Exploits of the Mixed IPID Assignment. In \textit{2020 ACM SIGSAC Conference on Computer and Communications Security (CCS'20), November 9-13, 2020, Virtual Event, USA.} ACM, NewYork, NY, USA, 13 pages. \url{https://doi.org/10.1145/3372297.3417884}}

\section{Introduction}
\label{sec:introduction}

Since the transmission control protocol (TCP) was first presented in RFC 793 in 1981~\cite{rfc793}, more than 100 TCP related RFCs have been released to improve the protocol~\cite{rfc7414}. Consequently, it becomes difficult for off-path attackers to hijack TCP connections, mainly due to the challenge of inferring the 32-bit random sequence numbers and acknowledgment numbers of a targeted TCP connection~\cite{rfc793, rfc4953, rfc6528,rfc6056}. When launching a brute-force attack, the attacker has to flood more than 300 million spoofed packets at a time to the target systems that support both RFC 793~\cite{rfc793} and RFC 5961~\cite{rfc5961}. Hence, off-path TCP attacks mainly rely on discovering side channel vulnerabilities to facilitate the inference of the sequence and acknowledgment numbers~\cite{gilad2014off,cao2018off,cao2016off,chen2018off,cao2019principled}.
Fortunately, most of the uncovered vulnerabilities have been fixed or constrained by the security community~\cite{gilad2014off,cao2018off,cao2016off}.

In this paper, we uncover a new off-path TCP hijacking attack that exploits the mixed IPID assignment method in the latest Linux kernels (i.e., version 4.18 and beyond) to either terminate victim TCP connections or inject malicious data into victim TCP connections.
First, our attack tricks the victim Linux machine into adopting the hash-based IPID assignment policy, instead of the by default more secure per-socket-based IPID assignment policy, on socket protocols such as TCP and UDP. Once the IPID assignment policy for socket protocols (TCP in our attack) is downgraded, it builds a side channel based on the IPID hash collisions of the globally shared 2048 hash counters, i.e., identifying a shared IPID counter on the victim by leveraging hash collisions. Second, by observing the shared IPID counter, an off-path attacker can detect the presence of TCP connections on the victim. Third, the attacker infers sequence and acknowledgment numbers of the victim connection to completely hijack the connection. This new attack does not need any assistance of puppets, i.e., unprivileged applications or sandboxed scripts controlled by attackers  on victim hosts~\cite{qian2012collaborative,qian2012off,gilad2014off}.

%
%

The \texttt{Identification} field of IP protocol (IPID) is used to indicate the uniqueness of a packet~\cite{RFC791,rfc6864}. After abandoning two previous vulnerable IPID assignment methods (i.e., global IPID assignment and per-destination IPID assignment)~\cite{ensafi2010idle,ensafi2014detecting,gilad2014off,IPID-perdes,Knockelcounting}, Linux currently assigns IPID to packets based on a mixed method~\cite{Linux,alexander2019detecting,zhang2018onis}. 
If a packet is generated from socket protocols such as TCP and UDP, Linux uses the per-socket-based IPID assignment policy that assigns IPID to the packet based on the counter recorded in the protocol socket. Otherwise, Linux adopts the hash-based IPID assignment policy that assigns IPID based on one of the 2048 globally shared hash counters. Since the counter recorded in the protocol socket cannot be observed by off-path attackers, the per-socket-based IPID assignment is more secure against off-path attacks. 
Linux uses the \texttt{DF} (Don't Fragment) flag in the packet header~\cite{RFC791} to choose between the two policies, since only socket protocols can set this flag to \texttt{TRUE} to perform the path MTU discovery (PMTUD) mechanism~\cite{rfc1191,rfc1981}. In other words, if the \texttt{DF} flag is set to \texttt{TRUE}, it uses the per-socket-based policy; otherwise, it chooses the hash-based policy. 
However, our study shows that the mixed IPID assignment in Linux implementations has vulnerabilities that can be exploited to  launch a new off-path TCP hijacking attack.

Since Linux uses the more secure per-socket-based IPID assignment by default for TCP connections, an off-path attacker first tricks the victim into assigning IPID for its TCP packets using the less secure hash-based IPID assignment. 
This goal can be achieved by pretending to be a router and sending a forged ICMP  ``Fragmentation Needed'' error message~\cite{rfc792} to a victim. Since the ICMP error message informs the victim that the packets issued from the victim need to be fragmented and the \texttt{DF} flag is set,  the victim will be tricked into cleaning the \texttt{DF} flag of TCP packets and thus uses the hash-based IPID assignment. Next, the victim chooses one IPID counter from the 2048 hash counters to assign IPID for its TCP packets. Among the 2048 globally shared hash counters, the target counter is decided by the hash value of four components, i.e., three fields of the packet (\textit{source IP address}, \textit{destination IP address}, \textit{protocol number}) and a \textit{random value} generated on system boot. Due to the small-sized hash counter pool, the attacker may identify the target hash counter used in a victim TCP connection via hash collisions, namely, alternating IP addresses to collide with the target counter.


%



Once the shared IPID counter is known, attackers can use the challenge \texttt{ACK} mechanism~\cite{rfc5961} as trigger conditions to change the shared IPID counter, facilitating the next two attack steps, i.e., to detect the presence of the victim TCP connection and infer the sequence and acknowledgment numbers. The attacker sends forged TCP packets to the victim, and the triggered challenge \texttt{ACK} packets will alter the shared IPID counter under different situations. It helps the attacker to determine if the specified values in the forged TCP packets are correct. Note that our attack only leverages the challenge \texttt{ACK} mechanism as trigger conditions to assist the inference of a victim TCP connection, instead of directly exploiting vulnerabilities in the challenge \texttt{ACK} mechanism to hijack TCP connections~\cite{cao2016off,cao2018off}.
%

Our attack does not suffer from traditional noise challenges that other works have to address~\cite{ensafi2010idle,ensafi2014detecting,pearce2017augur,pearce2018toward}. Since, in our attack, irrelevant TCP traffic using per-socket-based counters, instead of the hash-based counters, will not interfere with the attack traffic. Moreover, we measure that non-TCP traffic also rarely interferes with the attack. We evaluate the impacts of the new off-path TCP vulnerability on the Internet. We find that more than 20\% of the Alexa (www.alexa.com) top 100k websites are vulnerable to our off-path attack. Those websites can be tricked into cleaning the \texttt{DF} flag and downgrading the IPID assignment from the per-socket-based policy to the hash-based policy for their TCP packets after receiving forged ICMP ``Fragmentation Needed'' messages. We implement a PoC and perform case studies on a wide range of applications, e.g., HTTP, SSH and BGP, to validate the effectiveness of the attack. For example, an off-path attacker can detect and tear down a SSH connection in 155 seconds on average and manipulate web applications or BGP routing tables within 215 seconds. The average success rate of our exploit is over 88\%. These results demonstrate that the off-path TCP exploit could cause serious damages in real world. 

Finally, we propose countermeasures that aim to eliminate the root cause of the newly discovered off-path TCP attack. We fix the mixed IPID assignment in Linux kernels by determining if a packet is originated from TCP protocol on the \texttt{Protocol} field in IP header, instead of the \texttt{DF} flag. We implement a prototype of our countermeasure in Linux 4.18 and confirm its effectiveness through experimental evaluation on the Internet.

\noindent \textbf{Contributions}. Our main contributions are the following:
\begin{itemize}[leftmargin=*]
\setlength{\itemsep}{0.2pt}
	\item We uncover that the new mixed IPID assignment method can still be exploited to hijack TCP connections by  off-path attackers. 

	\item We uncover a new side channel in IPID assignment in the latest Linux kernels. We demonstrate that the side channel can be exploited to learn the presence of victim TCP connections and infer the sequence and acknowledgment numbers of the connections. 
	
	\item We measure the Alexa top 100k websites and find that more than 20\% of them are vulnerable to our off-path attack. We also perform case studies on a wide range of applications on the Internet and confirm the effectiveness of the attack.
	
	\item We analyze the root cause of the new attack and develop countermeasures that use new IPID assignment methods for TCP packets. Our prototype in Linux 4.18 validates its effectiveness.
\end{itemize}

\section{Background}
\label{sec:background}
In this section, we first introduce the IPID assignment policies adopted in the latest Linux kernels. Next, we describe two key mechanisms in TCP/IP operations, i.e., path MTU discovery and challenge \texttt{ACK}, which are related to develop our exploit.

\subsection{IPID Assignment in  Linux}\label{IPID_assign}
\begin{figure}[h]
	\vspace{-1mm}
	\begin{center}
		\includegraphics[width=0.45\textwidth, height=1.4in]{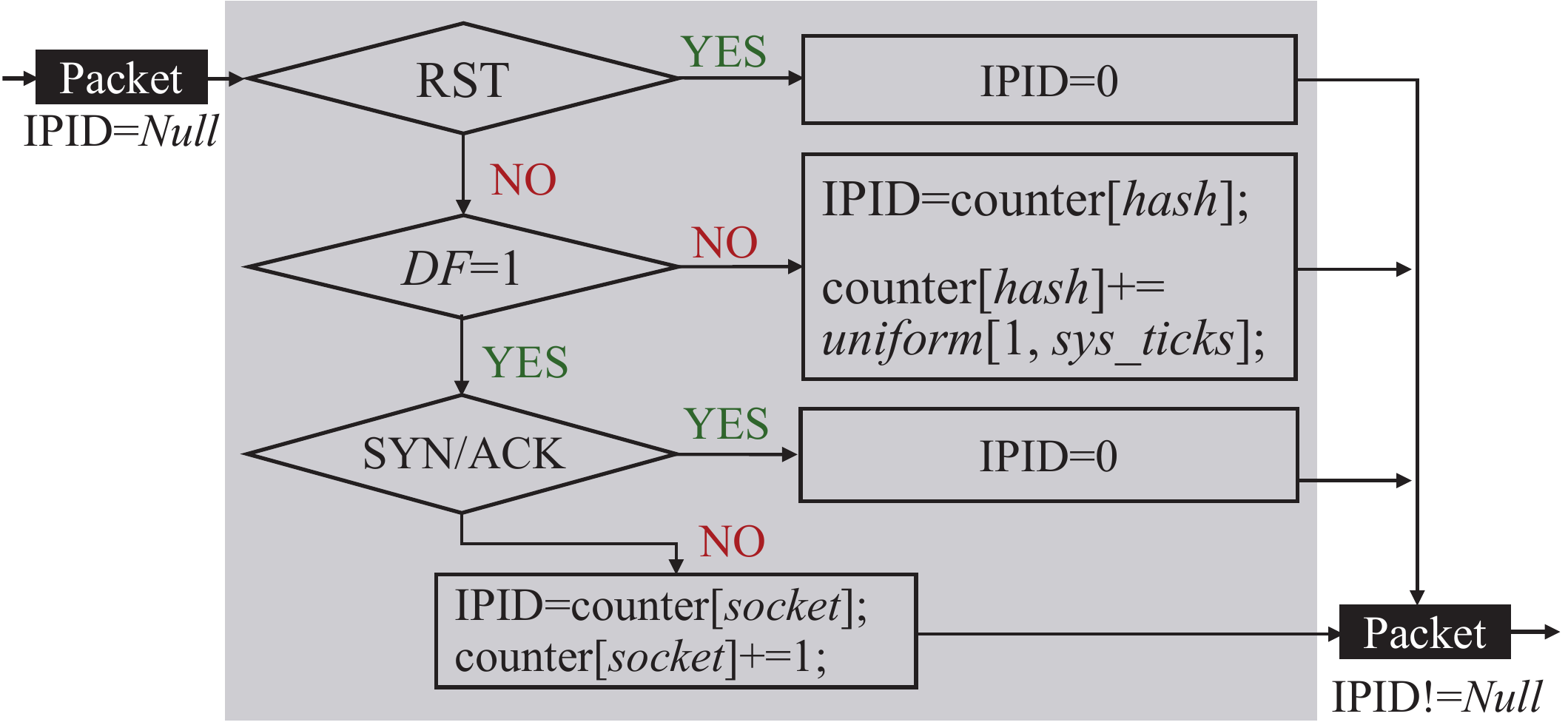}
		\vspace{-2mm}
		\caption{IPID assignment in Linux version 4.18 and beyond.}
		\label{assignment}
	\end{center}
	\vspace{-3mm}
\end{figure}

There are two basic IPID assignment policies in current Linux, i.e., IPID based on 2048 hash counters or IPID based on per-socket counters, where the latter is specific to socket related protocols such as TCP. Figure \ref{assignment} illustrates the procedure of IPID assignment in Linux version 4.18 and beyond. When a packet is generated, the IP protocol first checks whether the packet is a TCP \texttt{RST} packet. If yes, then the IPID of the packet is set to 0 directly. This assignment is due to Geoffrey \textit{et al.}'s disclosure of a side channel in previous assignment methods, i.e., IPID of the \texttt{RST} packet was assigned based on one of the 2048 hash counters before version 4.18, which can be exploited to detect the presence of TCP connections~\cite{alexander2019detecting}. 

If the packet is not a TCP \texttt{RST} packet, IP protocol checks the \texttt{DF} flag of the packet. IF the \texttt{DF} flag is set to \texttt{FALSE}, the IPID will be assigned based on a hash counter. There are totally 2048 hash counters in Linux. IP will select one from these counters according to the hash value of 4 variables, i.e., \textit{source IP address} and \textit{destination IP address} of the packet, the \textit{protocol number} of the packet, and a \textit{random value} generated on system boot. After the IPID value is copied from the selected counter, the counter will increase by a uniform distribution value between 1 and the number of system ticks since the last packet transmission that used the same counter. The system tick is usually measured in milliseconds.

When the \texttt{DF} flag is set to \texttt{TRUE}, except for the TCP \texttt{SYN/ACK} (both the \texttt{SYN} flag and the \texttt{ACK} flag in TCP header are set to \texttt{TRUE}) packet whose IPID is assigned to 0, IP assigns IPID for other packets based on the second policy that is based on a per-socket counter unique to each connection. The per-socket counter is usually initialized to a random value. Then each time after a packet is transmitted using this counter, the counter increases by 1. The per-socket-based assignment policy is considered to be more secure and can avoid being observed  from off-path attackers. Since the \texttt{DF} flag of TCP packets is be default set to \texttt{TRUE} to enable the path MTU discovery mechanism, TCP packets follow this per-socket assignment policy. 

Through studying the IPID assignment in Linux, we find that if the \texttt{DF} flag of TCP packets can be cleared (i.e., set to \texttt{FALSE}), then the IPID assignment to TCP packets will be downgraded from using the per-socket-based policy to the hash-based policy.

\subsection{Path MTU Discovery}\label{sec-PMTUD}
To avoid IP fragmentation, RFC 1191~\cite{rfc1191} and RFC 1981~\cite{rfc1981} propose a mechanism to discover path MTU (PMTU) between two end hosts, i.e., the minimum of all hops' MTUs in the entire packet transmission path. PMTUD relies on the \texttt{DF} flag. Before sending a packet, the originator sets the \texttt{DF} flag of the packet to \texttt{TRUE}, indicating that the packet is not allowed to be fragmented by intermediate routers. If the packet exceeds a router's next-hop MTU, the intermediate router discards it and issues an \textit{ICMP Destination Unreachable} message (type 3) to the originator with the code \textit{Fragmentation Needed and DF set} (code 4) in IPv4 or an ICMPv6 \textit{Packet Too Big} message in IPv6, along with the router's next-hop MTU value carried in the ICMP message. After receiving the ICMP message, if the embedded packet in the message passes the originator's check, then the originator reduces the size of subsequent packets according to the carried next-hop MTU value in the message. The originator repeats the sending process until a packet with certain size could be forwarded to the destination, and it then sets the size as PMTU.

However, during this procedure, if an intermediate router's next-hop MTU is smaller than the originator's acceptable minimum PMTU \textit{min\_pmtu} that is a system variable in PMTUD implementations, the originator will resize the packet size to \textit{min\_pmtu},  clear the \texttt{DF} flag of subsequent packets, and then send them out. In RFC 1191~\cite{rfc1191}, \textit{min\_pmtu} is recommended as 576 octets. However, it varies in different implementations, e.g., 256 octets in FreeBSD, 296 octets in Mac OS, 552 octets in Linux, and 596 octets in Windows.

In most PMTUD implementations, hosts do not validate the source and transmission path of ICMP ``Fragmentation Needed'' messages (e.g., Linux kernel version 3.9 and beyond). Therefore, an off-path attacker can pretend to be a router and forge such an ICMP message specified with an extremely small next-hop MTU value. Actually the specified next-hop MTU value can be even set to 68 octets, the minimum of PMTU value on the Internet. After sending such a forged ICMP message to the originator, if the embedded packet in the forged ICMP message can pass the originator's check, the originator will be tricked into clearing the \texttt{DF} flag, thus downgrading the IPID assignment for TCP packets. According to RFC 792~\cite{rfc792}, the forged ICMP message should embed at least 28 octets data to pass the originator's check. We will show that an ICMP echo reply packet can be embedded in the forged ICMP message to deceive the originator's check. 

\subsection{Challenge \texttt{ACK} Mechanism}\label{challenge-ack}
To defeat blind in-window attacks on TCP, the challenge \texttt{ACK} mechanism was proposed as RFC 5961~\cite{rfc5961}. In a nutshell, the challenge \texttt{ACK} mechanism requires that the sender of packets triggering the challenge conditions replies with the exact sequence number, not just one within the receive window.  Thus, it can prevent an off-path attacker's blind injection unless the attacker is extremely lucky to be able to guess the exact sequence number with a probability of $1/2^{32}$. The challenge \texttt{ACK} mechanism is designed to enhance the security of TCP; however, we show that it can be abused to infer the state of a victim TCP connection. 

Our attack exploits the challenge conditions in three aspects.
First, if a receiver sees an incoming \texttt{SYN} segment, regardless of the sequence number in the segment, it sends back an challenge \texttt{ACK} to the sender to confirm the loss of the previous connection. Only the legitimate remote peer will send a \texttt{RST} segment with the correct sequence number (derived from the \texttt{ACK} field of the challenge \texttt{ACK} packet) to prove that the previous connection is indeed terminated. Off-path attackers cannot answer this challenge with correct sequence number. We will show that this challenge condition can be abused to detect victim TCP connections. 

Second, when a receiver sees an incoming \texttt{RST} segment, if the sequence number of the segment is outside the receive window, the receiver simply discards the segment. Instead, if the sequence number is in-window but does not exactly match the expected next sequence number (i.e., $RCV.NXT$), the receiver will send a challenge \texttt{ACK} to the sender to confirm the reset action. We will show that this challenge condition can be abused to judge the correctness of the guessed sequence number. 

Third, if a receiver sees an incoming \texttt{ACK} segment, it validates the acknowledgment number of the segment ($SEG.ACK$) with a window of $SND.UNA - SND.MAX.WND <= SEG.ACK <= SND.NXT$, where $SND.UNA$ is the sequence number of the first unacknowledged octet, $SND.MAX.WND$ is the maximum window size that the receiver has ever seen from its peer. The receiver considers that the acknowledgment number is legal and accepts it  if the acknowledgment number is in this range. If $SEG.ACK$ is in the range of [$SND.UNA - (2^{31}-1)$, $SND.UNA - SND.MAX.WND$], i.e., the challenge \texttt{ACK} window, the receiver responds with a challenge \texttt{ACK} packet. We will show this challenge condition can be abused to judge the correctness of guessed acknowledgment number.

\section{Attack Overview}
\label{sec:overview}

\subsection{Threat Model}
\begin{figure}[h]
	\vspace{-2mm}
	\begin{center}
		\includegraphics[width=0.32\textwidth, height=0.8in]{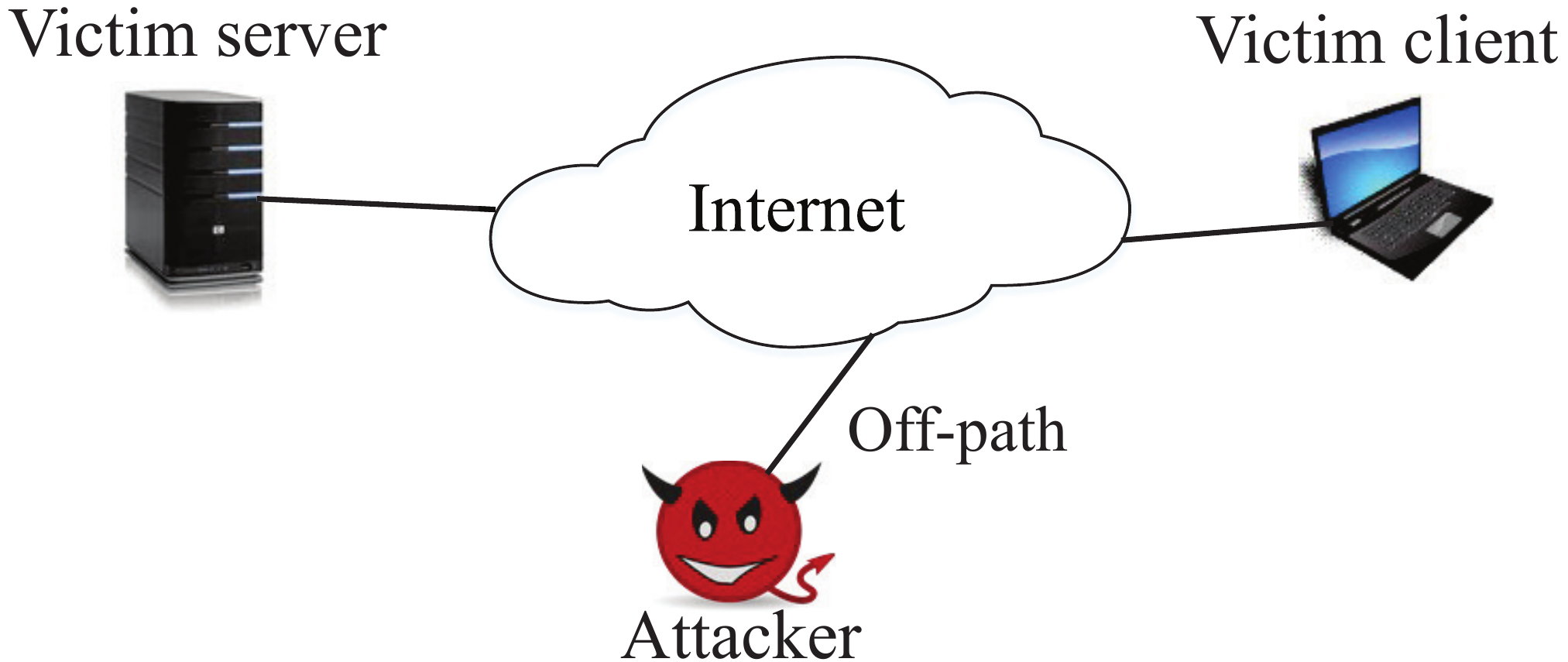}
		\vspace{-2mm}
		\caption{Threat model}
		\label{threat-model}
	\end{center}
	\vspace{-5mm}
\end{figure}

Figure \ref{threat-model} illustrates the threat model of our off-path TCP exploit. It involves three hosts, i.e., a victim client, a victim server, and an off-path attacker. The server and the client communicate based on a TCP connection, while the off-path attacker aims to hijack the connection.
The off-path attacker cannot eavesdrop the traffic transferred between the server and the client as the man-in-the-middle attacker does. However, the attacker is capable of sending spoofed packets with the IP addresses of the server and the client. This capability assumption is practical, since at least a quarter of the Autonomous Systems (ASes) on the Internet do not filter packets with spoofed source addresses leaving their networks~\cite{luckie2019network}.  

\subsection{Attack Procedure}
Our off-path TCP exploit consists of three main steps to hijack a victim TCP connection.

\noindent \textbf{Step 1: Detecting Victim Clients.} The attacker first downgrades the server's IPID assignment from the per-socket-based policy to the hash-based policy. Then, through hash collisions, the attacker detects victim clients who share the same IPID counter with the attacker on the server side, i.e., the server uses the same hash-based IPID counter to assign IPID for TCP packets to the victim client and for packets to the attack machine.

\noindent \textbf{Step 2: Detecting TCP Connections.} Once a potential victim client is detected, the attacker impersonates the victim client and sends spoofed \texttt{SYN/ACK} packets to the server. Then, by observing the change of the shared IPID counter, the attacker can determine the correctness of the specified source port number in the spoofed \texttt{SYN/ACK} packets and thus detect the presence of the TCP connection between the server and the victim client. 

\noindent \textbf{Step 3: Inferring Sequence and Acknowledgment Numbers.} After a victim TCP connection is identified, the attacker sends spoofed \texttt{RST} packets and \texttt{ACK} packets to the connection, and triggers challenge \texttt{ACK} mechanism on the connection. By observing the changes of the shared IPID counter, the attacker can determine the correctness of the specified sequence number and acknowledgment number in the forged packets.

After correctly identifying the sequence numbers and acknowledgment numbers of the victim connection, the attacker can forge malicious TCP segments with the identified values and inject the segments into the victim connection to either reset the connection or poison the data stream. In the next three sections, we will detail the above three steps.

\section{Detecting Victim Clients}
\label{sec:side-channel}
In this section, we present the method of building the IPID side channel to detect potential victim clients. First, we present the method of downgrading the IPID assignment from the  per-socket-based policy to the hash-based policy. Second, we describe how to detect potential victim clients who share the same IPID counter with the attacker on the server side.

\subsection{Downgrading the IPID Assignment}
Linux assigns IPID for packets based on the \texttt{DF} flag. If the \texttt{DF} flag is set to \texttt{TRUE}, Linux will assign IPID for the packet based on a per-socket IPID counter;  otherwise, based on a hash IPID counter. However, we observe that the \texttt{DF} flag can be maliciously cleared by off-path attackers, thus downgrading the IPID assignment. The attacker pretends to be a router and sends a forged ICMP ``Fragmentation Needed'' message to the victim server, indicating that a router between the server and the client has a smaller next-hop MTU and the packet is not allowed to be fragmented. 

In order to trick the server into accepting the forged ICMP ``Fragmentation Needed'' message and clearing the \texttt{DF} flag of TCP packets sent to the client, the forged ICMP message needs to satisfy two conditions. First, the server does not validate the source of the ICMP message, i.e., the forged ICMP message from off-path attackers will not be discarded by the server. In practice, the validation requires extra functionality support from hardware devices~\cite{wu2018enabling}, since major OSes, e.g., Linux 3.9 and beyond, do not perform the validation but directly accept the message. Second, the data embedded in the forged ICMP message must be able to evade the server's checks. RFC 792~\cite{rfc792} states that ICMP error messages should be embedded at least 28 octets (i.e., the IP header plus at least the first 8 octets) of the triggering packet, which is used by the server to match the message to the appropriate process. Moreover, according to the newer standard RFC 1812~\cite{rfc1812}, ICMP error messages should be embedded as much of the triggering packet as possible, but not exceeding 576 octets. Hence, the attacker has to craft and embed feasible data into the forged ICMP error message to evade the server's check.

\begin{figure}[h]
	\vspace{-2mm}
	\begin{center}
		\includegraphics[width=0.49\textwidth]{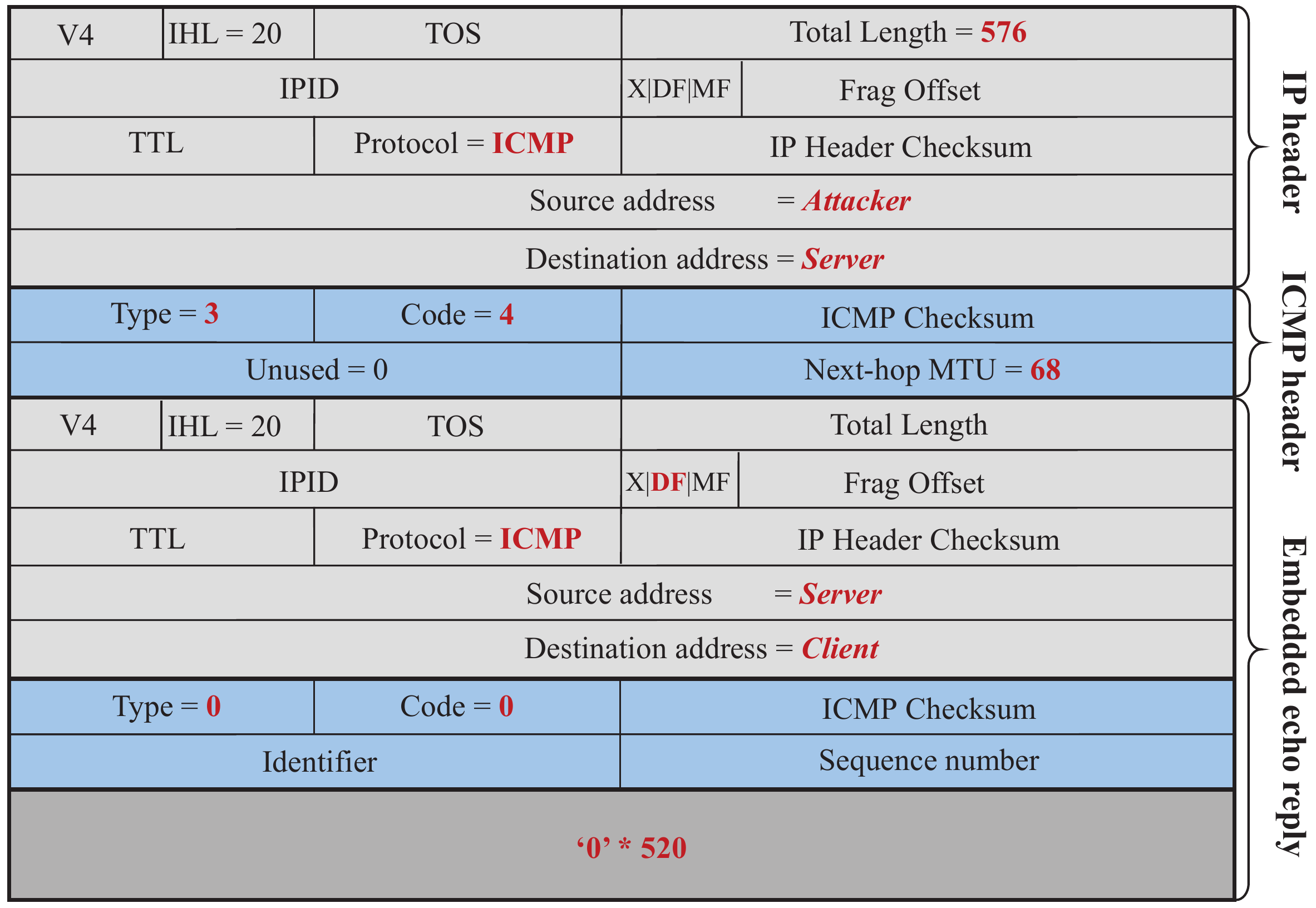}
		\vspace{-4mm}
		\caption{Structure of the forged ICMP error message.}
		\label{fragmentation_pic}
	\end{center}
	\vspace{-4mm}
\end{figure}

\begin{figure*}[h]
\vspace{-2mm}
	\begin{center}
		\subfigure[No hash collisions with the client]{ 
			\label{ack_no}  
			\includegraphics[width=0.488\textwidth]{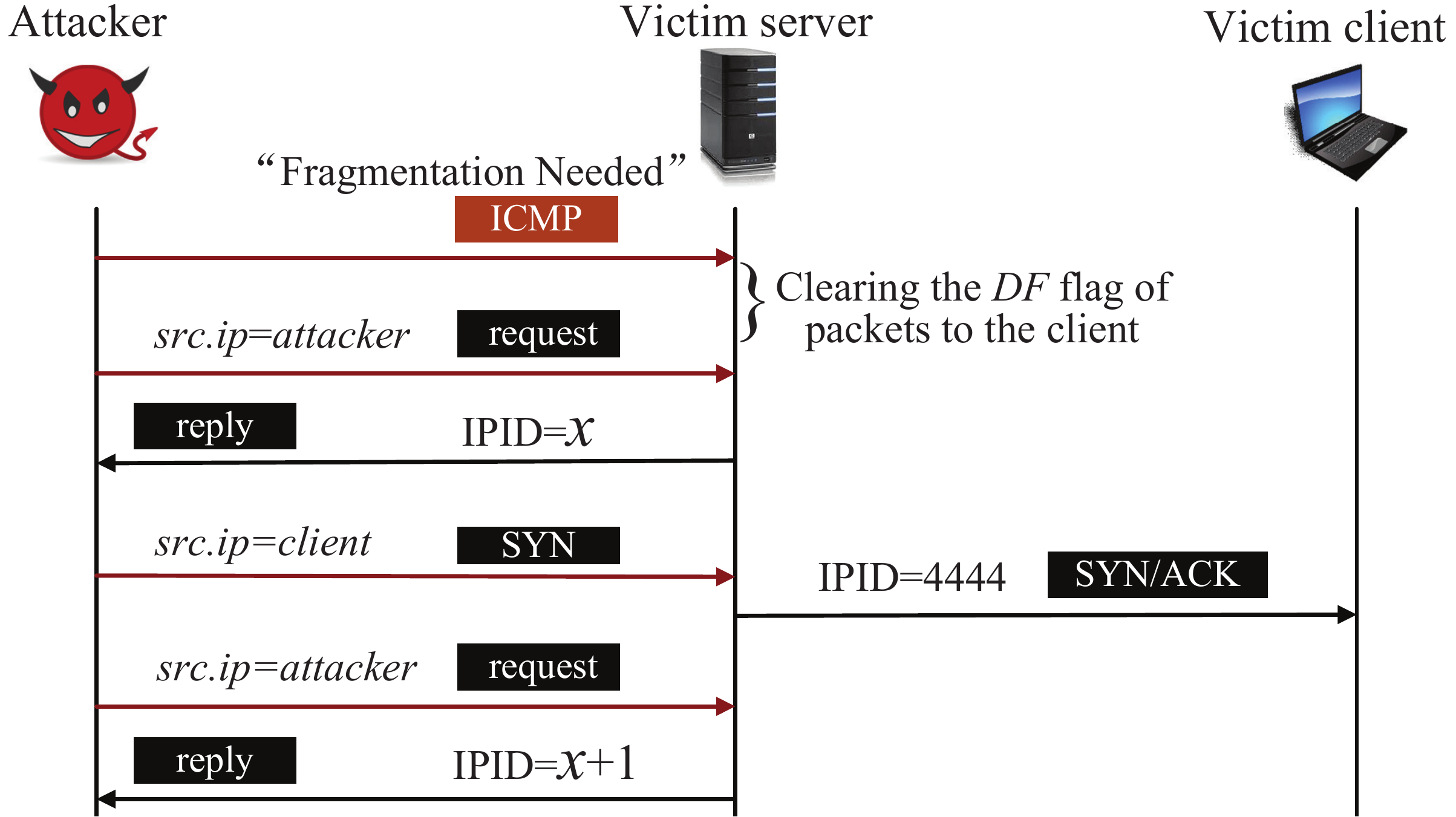} 
		} 
		\subfigure[Hash collisions with the client]{ 
			\label{ack_yes}
			\includegraphics[width=0.488\textwidth]{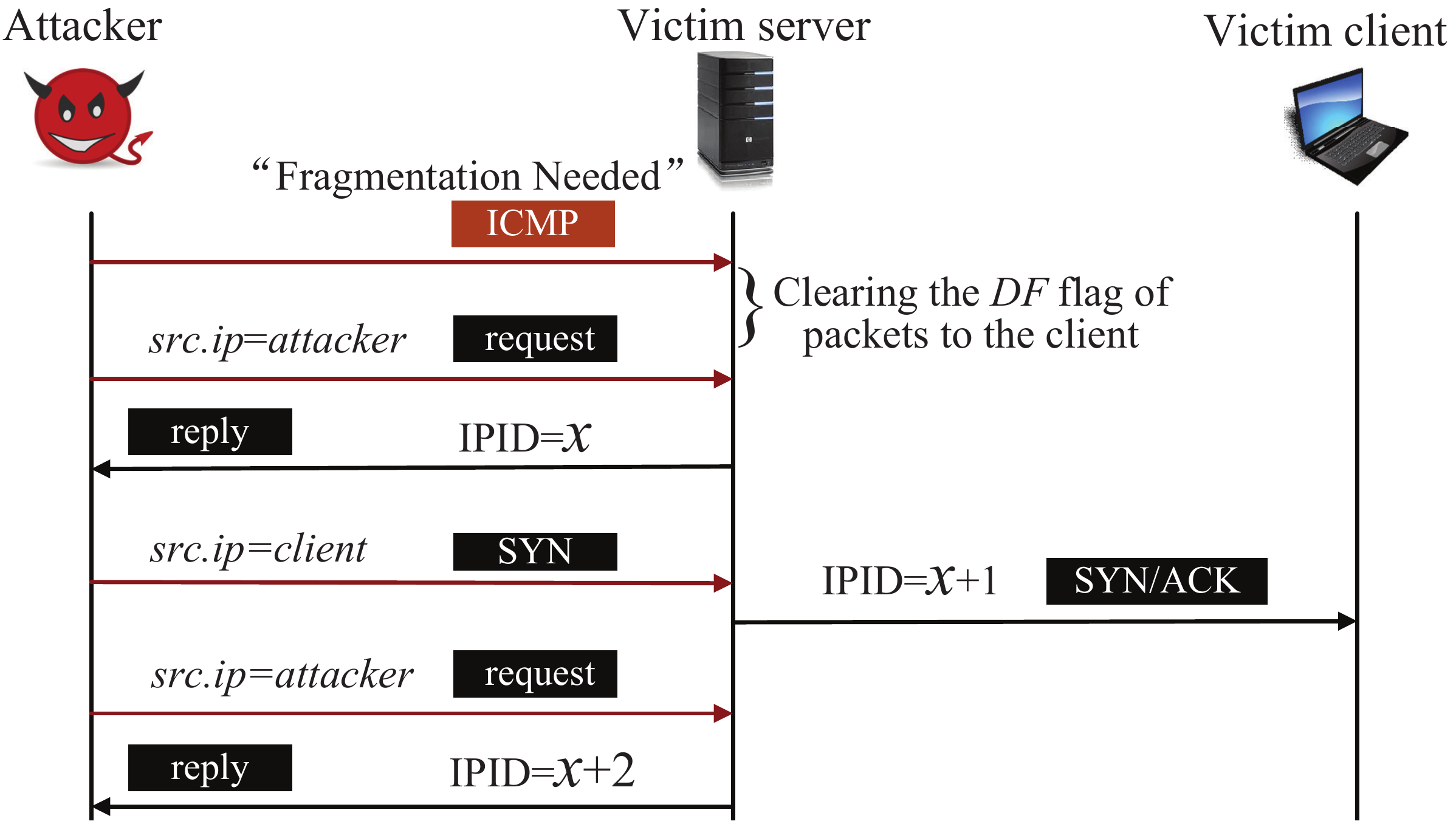}
		}
		\vspace{-4mm}
		\caption{Detecting potential victim clients through hash collisions.} 
		\label{discover_collision} 
	\end{center}
	\vspace{-5mm}
\end{figure*}

To evade the server's check, we can embed the ICMP echo reply data into the forged ICMP ``Fragmentation Needed'' message, as shown in Figure~\ref{fragmentation_pic}. When servers equipped with Linux 3.9 and beyond receive such an ICMP ``Fragmentation Needed'' message embedded with an echo reply, the server does not check whether it sent the embedded echo reply data earlier. Instead, it directly responds to the forged ICMP error message and clears the \texttt{DF} flag of subsequent packets sent to the client whose IP address is specified in the embedded echo reply. Even if the server checks on the embedded echo reply data, it is easy for attackers to circumvent this checking. 
For example, the attacker can impersonate the client and initiate an ICMP echo request to the server, triggering the server to send an echo reply message. Then, the attacker pretends to be a router and sends an ICMP ``Fragmentation Needed'' message embedded with the known echo reply data to the server, tricking the server into accepting the forged message. 
Note the next-hop MTU value specified in the forged ICMP ``Fragmentation Needed'' message should be smaller than the server's acceptable minimum PMTU \textit{min\_pmtu}, as described in Section~\ref{sec-PMTUD}. Actually, the value can be set to 68 octets, which is always smaller than the system variable of \textit{min\_pmtu} in various IP implementations. Besides, we find that a forged ICMP error message embedded with a GRE data~\cite{rfc2784} can also be used to trick the server into clearing the \texttt{DF} flag.

In a nutshell, it is difficult to verify the legitimacy of ICMP error messages on the Internet. Therefore, an attacker can forge an ICMP message and trick the server into accepting it. The forged ICMP error messages can force the server to clear the \textit{DF} flag of packets sent to the victim client. Thus, the IPID assignment can be easily downgraded by attackers.

\subsection{Constructing Hash Collisions}\label{hash-collsions}
The server will assign IPID to the packets by using one of 2048 hash counters once the TCP packet's \texttt{DF} flag is cleared. The counter is selected based on four factors, i.e., \textit{source IP address}, \textit{destination IP address}, \textit{protocol number} (e.g., 1 represents ICMP protocol, 6 represents TCP protocol) of the packet, and a \textit{random value} generated on system boot. A hash index computed from the four factors is used to select one counter from the 2048 hash IPID counters. Hence, if the TCP packets sent to the victim client have the same hash value as the packets sent to the attacker, the server will use the same IPID counter to assign IPID for those packets to different destinations.
Therefore, by constructing hash collisions using Equation~\ref{eq1}, the attacker can detect the victim clients who share the same IPID counter on the server side. In other words, the shared IPID counter forms a side channel, which can be exploited to infer TCP connections between the server and the detected client. Our attack uses the ICMP protocol to detect hash collisions due to its simplicity and observability.

\vspace{-2mm}
\begin{equation}
\begin{aligned}
&hash(server\_IP, \textit{\textbf{client\_IP}}, TCP, Boot\_key) = \\
&hash(server\_IP, \textit{\textbf{attacker\_IP}}, ICMP, Boot\_key)\label{eq1}
\end{aligned}
\end{equation}

The procedure of detecting victim clients by constructing hash collisions is shown in Figure~\ref{discover_collision}. First, the attacker pretends to be a router and sends a forged ICMP ``Fragmentation Needed'' message to the server, who will be tricked into clearing the \textit{DF} flag and downgrading the IPID assignment of packets to the client whose IP address is specified in the forged ICMP message. Second, the attacker initiates ICMP echo requests to the server and observes the IPID of the reply packets. Then the attacker impersonates the client and sends a spoofed \texttt{SYN} packet to the server's listening port (e.g., 80). Following the three-way handshake process of TCP, the server will respond an \texttt{SYN/ACK} packet to the client. Here, the key difference is that if the attacker's IP address collides with the client (i.e., the attacker and the client share the same hash-based IPID counter at the server side), the IPID assignment for the \texttt{SYN/ACK} packet will incur an additional increment to the shared IPID counter, which can be observed by the attacker \footnote{A  special case is that the source port in a spoofed \texttt{SYN} packet happens to match the source port of a TCP connection from the client to the server. In this case, the spoofed \texttt{SYN} packet will trigger a challenge \texttt{ACK} packet, instead of a \texttt{SYN/ACK} packet. However, the attacker can still observe an additional increment to the shared IPID counter.}. Otherwise, if there is no collision, the IPID observed by the attacker will be continuous distribution, i.e., without additional increment. Using this method, the attacker can identify victim clients who collide with its IP address and share the same IPID counter.

The hash-based IPID counter does not increase linearly. Instead, the increment is a random value in a uniform distribution between 1 and the number of system ticks since the last packet transmission that used the same counter. Hence, if the attacker wants to force the counter to increase linearly and facilitate the observation, it needs to restrict the increment of system ticks related to the IPID counter. Usually, if more than 3 packets are sent to the server under 10 $ms$, the random value added to the IPID counter will never be larger than one~\cite{alexander2019detecting}. We send ICMP request packets to the server in parallel and restrict the increment of system ticks. Our experiments show that if the round-trip time (RTT) from the attacker to the server is within 200 $ms$, the attacker only needs to send less than 300 packets per second to force the hash-based IPID counter increasing linearly.

There are totally 2048 hash-based IPID counters in Linux, and the probability of hash collisions between the attacker and the client is a geometric distribution. When the protocol is specified as ICMP (see Eq.~\ref{eq1}), if the attacker has \textit{k} IP addresses, the probability of collisions between the attacker and the target client is $1-(1-p)^{k}$, where $p$ equals 1/2048. To construct an attack in practice, the attacker has two strategies to detect victim clients by leveraging hash collisions. 

\noindent\textbf{Attacking Potential Targets.} If the attacker has only one or a few IP addresses, the attacker can detect potential victim clients who collide with the attacker. In theory, if the attacker has only one IP address, on a target server, the number of clients conflicting with the attacker is $2^{32}/2048 = 2^{21}$. Since the detection only depends on the server side, the attacker can create a list of IP addresses and select one from the list as the client's IP address each time. Following the procedure in Figure~\ref{discover_collision}, the attacker can determine if the selected one is a victim. In this way, the attacker can identify and enumerate all the potential victim clients who are vulnerable to its IP address. In our test, an attacker can detect more than 20 victim clients within 6 minutes using one IP address (see Section~\ref{conllison-dec}).

\noindent\textbf{Attacking Arbitrary Targets.} If the attacker has enough number of IP addresses, it can attack arbitrary TCP connections by alternating its IP addresses to generate the hash collision. According to the geometric probability distribution, if the attacker has more than 2048 IP addresses, it can collide with any clients with above 63.2\% probability. Especially, for servers having the IPv4 and IPv6 dual-stack, attackers can use IPv6 addresses to construct hash collisions with arbitrary target clients, since both IPv4 and IPv6 use the same 2048 hash-based IPID counters~\cite{zhang2018onis}.

The detected hash-based IPID counter shared with the victim client is stable. That is, if the server does not restart (i.e., the $Boot\_key$ in Eq.~\ref{eq1} is not altered), the client's TCP connection will always share this IPID counter with the attacker. In practice, servers (e.g., web servers and BGP routers) do not restart frequently. Hence, the attacker can detect shared IPID counters and victim clients in advance, regardless of if TCP connections have been established.

Note Linux assigned IPID to \texttt{RST} packets based on hash counters before version 4.18, and thus an attacker can observe its IPID distribution to determine if it shares the same counter with the client by spoofing \texttt{SYN/ACK} packets~\cite{alexander2019detecting}. This vulnerability has been fixed since Linux 4.18 by always setting the IPID of \texttt{RST} packets  to 0, incurring no changes on any IPID counters. However, we find that after the critical step of downgrading the IPID assignment, an attacker can still detect hash collisions through forging SYN packets and triggering the server to respond \texttt{SYN/ACK} packets and then identify a victim client. Moreover, we will show that the fix of assigning 0 to \texttt{RST} packets introduces yet another vulnerability, which can be exploited by a pure off-path attacker to detect the presence of victim TCP connections (see Section~\ref{sec:connection-detec}).

\section{Detecting TCP Connections}
\label{sec:Connection}
Once a victim client is identified, the attacker can learn the presence of TCP connections between the client and the server. 


\subsection{TCP Connection Detection}\label{sec:connection-detec}
A TCP connection is identified by a four-tuple, i.e., [source IP address, source port number, destination IP address, destination port number]. Usually, the destination IP address, and port number are public known, so an attacker only needs to infer the source IP address and source port number. In our attack, since the victim client can be detected by using hash collisions, the only missing tuple is the source port number.

Assuming that a TCP connection from source port \textit{y} has been established earlier by a legal user in the victim client, an off-path attacker can identify this port number by sending out probing packets. First, the attacker continuously sends ICMP echo request packets to the server and observes the IPID values of the reply packets from the server. Then, the attacker impersonates the victim client and sends a forged \texttt{SYN/ACK} packet with a guessed source port number to the server. If the source port number specified in the \texttt{SYN/ACK} packet does not equal \textit{y}, according to the TCP specification~\cite{rfc793}, the server will respond a \texttt{RST} packet to the client. Due to the patch fixing the vulnerability identified by Alexander \etal~\cite{alexander2019detecting}, Linux kernel versions 4.18 and beyond assign an IPID of 0 to the \texttt{RST} packet, which will not incur an increment to the shared IPID counter. Hence, the IPID values in the reply packets observed by the attacker are continuous.

If the guessed source port number specified in the forged \texttt{SYN/ACK} packet equals \textit{y}, the challenge \texttt{ACK} mechanism~\cite{rfc5961} makes the server send a challenge \texttt{ACK} packet to the victim client for confirming the legitimacy of the \texttt{SYN/ACK} packet. The IPID in the challenge \texttt{ACK} packet will be assigned based on the shared IPID counter, which will incur an additional increment to the counter. Thus, from the view of the attacker, the IPID values in the reply packets from the server will not be continuous. 

The attacker repeats the above procedure, i.e., changing the source port number specified in the forged \texttt{SYN/ACK} packet and then observing the IPID of the reply packets, until the correct port number \textit{y} is identified. Finally, the TCP connection running on the identified four-tuple is all known to the attacker. 

\subsection{Practical Considerations in Detection}\label{prctical}
There are three practical considerations that may have impact on the detection time cost and the success rate.

\noindent \textbf{(1) Unexpected Responses to \texttt{SYN/ACK} Packets.} According to the TCP specification, when receiving an unexpected \texttt{SYN/ACK} packet, the server responds a \texttt{RST} packet to the client. However, in practice, the \texttt{RST} packet may not be sent to the client in two circumstances. First, some TCP implementations may not follow the specification strictly, i.e., the server does not issue a \texttt{RST} packet when an unexpected \texttt{SYN/ACK} packet is received. Second, firewalls or other middleboxes may filter and discard the \texttt{RST} packet from the server. However, in either case, the detection of TCP connections will not be disturbed notably. The reason is that even if the \texttt{RST} packet is discarded or not generated, it is still the same as being accepted by the client and the shared IPID counter will not have an additional increment. Instead, when the source port number \textit{y} is identified, the challenge \texttt{ACK} packet will be certainly issued to the client due to the strict standard action enforced by RFC 5961, so the attacker will observe the change of the IPID counter. Therefore, the unexpected responses to \texttt{SYN/ACK} packets will not affect the detection of TCP connections.

\noindent \textbf{(2) Parallel Search for Source Port.} The maximum possible port range is $2^{16}$ (from 0 to 65535), but the default range of source port number on Linux is only from 32768 to 61000 and Windows has a more narrow source port range from 49152 to 65535\footnote{The OS types or versions of the client are unrestricted in our exploit.}. To facilitate the identification of the source port, the attacker can adopt a parallel approach to search the source port number by sending multiple probing packets in a certain range during a period. If the source port is in the range, the shared IPID counter will have an additional increment, so the attacker can further narrow the range. Otherwise, the attacker can detect another port range in parallel. In addition, the attacker can use a binary-search-like algorithm~\cite{cao2016off,cao2018off} to further reduce the detection time cost.

\noindent \textbf{(3) Rate Limit of Challenge \texttt{ACK}.} In order to avoid DoS attacks against the server, the implementations of challenge \texttt{ACK} usually enforce a rate limit to restrict the number of challenge \texttt{ACK} packets, e.g., the rate limit enforced on per TCP connection on Linux is no more than 1 challenge \texttt{ACK} packet per 500ms. Therefore, when detecting TCP connections using the parallel approach, we need to consider the rate limit of challenge \texttt{ACK}.  Assuming we have located the source port number in the range of [i, i + n], we need to further narrow the detection range and 
the next challenge \texttt{ACK} packet will be sent in 500 $ms$. In practice, the time cost is acceptable and will not have a notable impact on the detection time. In our test, we can detect a victim TCP connection within 40 seconds. 

\section{Inferring Sequence and Acknowledgment Numbers}
\label{sec:Numbers}
In this section, we first briefly review the checking mechanism for TCP sequence and acknowledgment numbers. To detect the exact sequence number and an acceptable acknowledgment number for successfully injecting a forged segment into the target TCP connection, we develop a four-step inference method. First, we infer acceptable sequence numbers in the server's receive window. Second,  based on the inferred sequence number, we locate the challenge \texttt{ACK} window. Third, we detect the lower boundary of the server's receive window (i.e., the exact sequence number) based on the identified acceptable sequence numbers and challenge \texttt{ACK} window. Finally, we detect the acceptable acknowledgment numbers via probing the boundary of the challenge \texttt{ACK} window and inferring the boundary of the server's send window. After obtaining those information, the attacker is able to inject malicious segments into the target TCP connection.

\subsection{Preliminaries: Verifying Sequence Number} 
An TCP segment receiver first checks the sequence number in the segment header based on its receive window when a TCP segment arrives, i.e., the condition of $RCV.NXT <= SEG.SEQ <= RCV.NXT + RCV.WND$ must be satisfied, where $SEG.SEQ$ is the sequence number of the received segment, $RCV.NXT$ is the sequence number of the next octet that the receiver expects to receive, and $RCV.WND$ is the receive window size. Besides, the receiver following RFC 5961 will check the acknowledgment number based on its acceptable \texttt{ACK} window as described in Section~\ref{challenge-ack}. After passing these two checks, the segment will be accepted. In current TCP implementations, the \texttt{ACK} flag is always set to \texttt{TRUE} except for the first \texttt{SYN} packet for establishing the connection. As a result, the guessing of the acknowledgment number cannot be circumvented by disabling the \texttt{ACK} flag.

Since TCP is a full duplex protocol, when both the sequence and acknowledgment number acceptable by one side are inferred, the attacker can also identify the sequence and the acknowledgment number acceptable by the other side. For example, in fact, the $RCV.NXT$ and $SND.NXT$ on the server are equivalent to $SND.NXT$ and $RCV.NXT$ on the client~\cite{rfc793,cao2016off,cao2018off}. Hence, the attacker only needs to infer the sequence and acknowledgment numbers of one side. In our attack, we focus on inferring the sequence and acknowledgment numbers acceptable by the server. 


\subsection{Inferring Acceptable Sequence Number}\label{approximate_seq}
To infer the acceptable sequence numbers on the server side, the attacker continuously sends ICMP request packets to the server and observes the IPID values of the reply packets. Then, the attacker impersonates the victim client to send a spoofed \texttt{RST} packet to the server. The \texttt{RST} packet is specified with the guessed sequence number \textit{seq}. We need to consider two cases: (i) \textit{seq} not in the server's receive window and (ii) \textit{seq} in the server's receive window. According to the challenge \texttt{ACK} mechanism described in Section~\ref{challenge-ack}, in the first case, the server will discard the spoofed \texttt{RST} packet directly, so it does not influence the shared IPID counter. In the second case when the guessed \textit{seq} is in the server's receive window, the server will respond to this \texttt{RST} packet and send a challenge \texttt{ACK} packet to the victim client to confirm the legitimacy of the packet. The IPID of this challenge \texttt{ACK} packet is assigned based on the shared IPID counter, and it will incur an additional increment to the counter. The attacker can observe the increment and then determine that the guessed \textit{seq} is located in the server's receive window\footnote{In a special case when \textit{seq} exactly matches the server's $RCV.NXT$, the server will reset the connection directly. However, the probability of this case occurring is $1/2^{32}$, which is negligible.}. 

In practice, in order to reduce the time cost of sequence number inference, the attacker can divide the sequence number space into multiple blocks whose sizes are equal to the default receive window size in Linux (87380 octets), probing only once per block. Besides, the attacker can apply parallel search methods similar to those used in connections detection to further reduce the time cost.

\subsection{Locating the Challenge \texttt{ACK} Window}\label{loc-ack-win}
According to RFC 5961, when a segment arrives at the server, the server also checks the segment's acknowledgment number even if its sequence number is in the server's receive window.
There are three cases in the whole acknowledgment number space: (i) the acknowledgment number in challenge \texttt{ACK} window, (ii) in the acceptable \texttt{ACK} range, and (iii) invalid acknowledgment numbers. In the first case, the server will issue a challenge \texttt{ACK} packet to confirm the legitimacy of the triggering segment. In the second case, the server will accept the segment directly. Otherwise, if the segment carries an invalid acknowledgment number, the server will discard it silently. The last two cases cannot be differentiated directly because it cannot be observed from an off-path attacker. However, the attacker can first identify the challenge \texttt{ACK} window of the server and then infer the acceptable \texttt{ACK} numbers. 

When locating the challenge ACK window, the attacker observes and records the shared IPID counter. Then the attacker impersonates the victim client and sends a spoofed \texttt{ACK} packet with a guessed acknowledgment number $ack$ to the server, the packet is also specified with an acceptable sequence number detected previously.
If $ack$ is in the challenge \texttt{ACK} window of the server, a challenge \texttt{ACK} packet will be issued, incurring an additional increment to the shared IPID counter. Instead, if $ack$ is not in the challenge \texttt{ACK} window, the observed IPID will be continuous from the view of the attacker. In practice, the challenge \texttt{ACK} window size is always between 1G and 2G~\cite{rfc7323, cao2016off, cao2018off}, i.e., one quarter of the entire acknowledgment number space. Hence, to facilitate the detection, the attacker can divide the entire space into 4 blocks and probe each block to check which block the challenge \texttt{ACK} window falls in.

\subsection{Detecting the Exact Sequence Number}\label{Exact_seq}
Now we present the method of detecting the exact sequence number (i.e., $RCV.NXT$, the lower boundary of the server's receive window) based on the previous inferred results. The attacker can forge multiple \texttt{ACK} packets with a constant acknowledgment number $ack\_challenge$ in the challenge \texttt{ACK} window and the specified sequence number in each \texttt{ACK} packet set to $seq\_acceptable - i$, where $seq\_acceptable$ is an acceptable sequence number inferred previously. Then the attacker impersonates the victim client to send these forged \texttt{ACK} packets to the server. In the beginning, the server will be triggered to send challenge \texttt{ACK} packets at a rate of one packet per 500 $ms$ due to the rate limit of challenge \texttt{ACK}, so the triggered challenge \texttt{ACK} packets will incur regular increments to the shared IPID counter. However, once the specified sequence number $seq\_acceptable - i$ reaches $RCV.NXT$ (the lower boundary of the server's receive window), the server will switch to send duplicate \texttt{ACK} packets, which is not enforced by any rate limit. Thus, the shared IPID counter will have a jitter\footnote{
The increments to the shared IPID counter become 20 per 500 $ms$ in our experiments, instead of 1 per 500 $ms$.}, and the attacker can observe this jitter and then detect the exact sequence number. The detecting procedure has no side effects, e.g., resetting the connections, on the connections.

\subsection{Detecting Acceptable \texttt{ACK} Number}\label{acceptable-ack}
Once an acknowledgment number $ack\_challenge$ in the challenge \texttt{ACK} window is identified, the attacker can also detect the boundary of the challenge \texttt{ACK} window by sending multiple probing \texttt{ACK} packets and then observing the shared IPID counter, similar to detecting the lower boundary of the server's receive window. The forged probing \texttt{ACK} packets are specified with a constant sequence number $seq\_acceptable$, and the acknowledgment number of each \texttt{ACK} packet is set to $ack\_challenge - i$. In turn, the attacker sends these forged \texttt{ACK} packets to the server. Challenge \texttt{ACK} packets will be triggered until $ack\_challenge - i$ reaches the lower boundary of the challenge \texttt{ACK} window. Once this boundary is detected, then $SND.UNA$ can be easily inferred, i.e., adding $2G$ to the detected boundary. $SND.UNA$ is in the acceptable \texttt{ACK} range. When all the data sent earlier has been acknowledged, $SND.UNA$ equals $SND.NXT$. Instead, if the server has an amount of data to be sent to the client, $SND.NXT$ can also be inferred by adding the typical size of the send window to $SND.UNA$, e.g., 16384 octets in Linux by default.

\section{Implementation and Evaluation}
\label{sec:Implementation}
We first conduct experiments to show feasibility of identifying victim clients on the Internet via hash collisions. Next, we conduct two case studies to evaluate the effectiveness and efficiency of our attacks. By launching the TCP connections DoS attacks, an off-path attacker can reset a TCP connection in 155 seconds on average. By conducting the TCP connections manipulation attacks, an off-path attacker can hijack the session and manipulate web traffic and BGP routing table within 215 seconds. 

\noindent \textbf{Ethical considerations.}
In order to avoid causing real damages or negative impacts on the Internet, we choose not to directly attack real users and their hosts. All the hosts involved in the experiments are our machines. We evaluate  the impacts of our off-path TCP attacks on the Internet, e.g., measuring Alexa top 100k websites to identify potential victim servers that are vulnerable to our attack. However, we do not exploit the vulnerability of these web servers for real attacks. 

\subsection{Identifying Victim Clients}\label{conllison-dec}
We show two scenarios of identifying victim clients via hash collisions after downgrading the server's IPID assignment. First, we show how to detect potential victim clients using one IP address. Second, we illustrate that the attacker can attack arbitrary victim clients if having enough IP addresses.

\begin{figure*}[t]
	\begin{center}
		\subfigure[Empirical CDF of time cost]{ 
		\label{colleage}
		\includegraphics[width=0.48\textwidth]{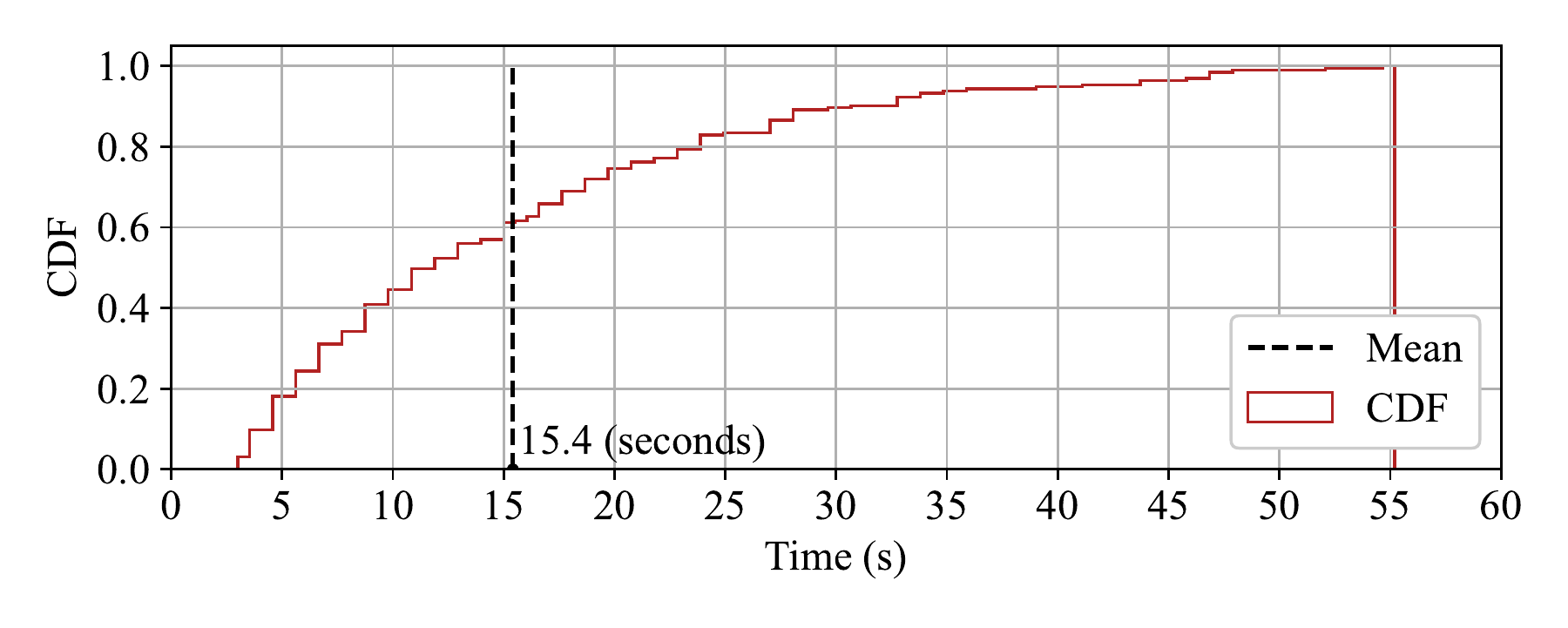}
	} 
	\vspace{-2mm}
	\subfigure[Empirical CDF of the number of attacker IP addresses]{ 
		\label{financial}
		\includegraphics[width=0.48\textwidth]{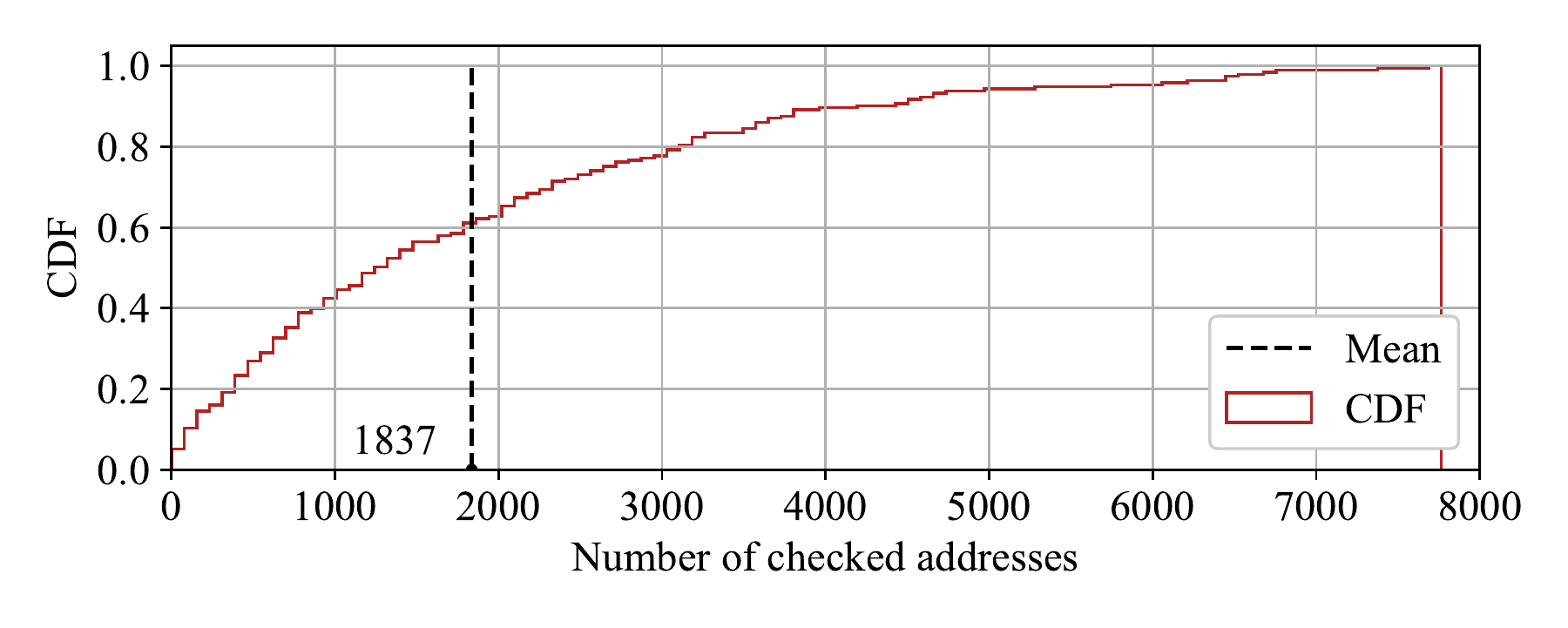}
	}
	\vspace{-2mm}
	\caption{Empirical CDF of time cost and the number of attacker IP addresses needed to detect a hash collision.} 
	\label{multi-scan-result}
	\end{center}
\vspace{-2mm}
\end{figure*}

\noindent\textbf{Experimental Setup.} Two types of hosts are used in this experiment. A server listening on port 80 and waiting for TCP connection requests is equipped with Ubuntu 18.04 (kernel version 5.5) with a prefix of 152.136.0.0/16. Attack machines locate in different positions with independent IP addresses. We use these IP addresses to detect victim clients to the server. The attack machines are equipped with Ubuntu 18.04 (kernel version 4.15) and are able to send packets to the server with spoofed IP addresses.

\noindent\textbf{Results with One Attacker IP Address.} 
When having only one IP address, an attacker can detect more than 2 million victim clients to the server. In this scenario, we deploy one attack machine and assign IP address from a target IP list, which contains several prefixes owned by different organizations. The attack machine clears the \texttt{DF} flag of the server's TCP packets to IP addresses in these prefixes via forging ICMP ``Fragmentation Needed'' messages, and then it scans the target prefixes to identify potential victim clients that share the same hash-based IPID counter with the attack machine on the server. The attacker terminates the scanning process after scanning all IP addresses within the prefix or reaching the time limit (30 minutes in our experiment). The experimental results are shown in Table~\ref{scan_results}. By using only one attack machine, the attacker can detect a considerable number of potential victim clients. For instance, the numbers of victim clients that collide with the attack machine are 179, 156, and 121 in the prefixes of 3.208.0.0/12, 101.80.0.0/12 and 50.16.0.0/14, respectively. It takes 14.0 seconds on average to detect a potential victim client. Average outbound traffic of the attack machine is 125.14 KB/s (i.e., around 584 packets/s). Thus, 
it is difficult to detect the malicious probing, e.g., by leveraging network traffic monitoring systems. 

\begin{table}[h]
	\caption{Detecting victim clients using one IP address.}
	\vspace{-2mm}
	\begin{center}
		\begin{tabular}{@{}r|llcc@{}} 
			\toprule
			\textbf{\begin{tabular}[c]{@{}c@{}}Prefix \\ Owner\end{tabular}} &
			\textbf{Prefix} & \textbf{Location} &
			\textbf{\begin{tabular}[c]{@{}c@{}}Victim \\ clients\end{tabular}} & \textbf{\begin{tabular}[c]{@{}c@{}}Time \\ (min)\end{tabular}} \\ \midrule
			eBay          & 209.140.128.0/18 & US, CA   & 4   & 1.32 \\ 
			Yahoo Japan   & 124.83.128.0/17  & JP, TKY  & 9   & 2.70 \\
			Google        & 74.125.0.0/16    & US, CA   & 23  & 5.56 \\
			Tencent       & 162.14.0.0/16    & CH, BJ   & 27  & 5.54 \\
			Facebook      & 157.240.0.0/16   & US, CA   & 33  & 5.53 \\
			Alibaba       & 47.56.0.0/15     & CH, HK   & 57  & 11.3 \\
			Amazon        & 50.16.0.0/14     & US, WA   & 121 & 22.9 \\
			China Telecom & 101.80.0.0/12    & CH, SH   & 156 & 30.0 \\
			Amazon        & 3.208.0.0/12     & USA, VA   & 179 & 30.0 \\ 
			\bottomrule
		\end{tabular}
		\label{scan_results}
	\end{center}
	\vspace{-4mm}
\end{table}

\noindent\textbf{Results with Multiple Attacker IP Addresses.} When the attacker has multiple IP addresses, it aims to attack an arbitrary client to the server using these IP addresses. After selecting the target client, the attacker clears the \texttt{DF} flag of the server's TCP packets to the client by forging an ICMP ``Fragmentation Needed'' message. Then, the attacker detects addresses in its address pool to find a correct one that collides with the target client, i.e., sharing the same hash-based IPID counter. In this experiment, we select different target clients and repeat the detecting process 200 times. The empirical cumulative distribution function (CDF) of the time cost and the number of required attacker IP addresses are shown in Figure~\ref{multi-scan-result}. For an arbitrary target client, the average time cost to detect a correct IP address in the attacker's address pool is 15.4 seconds, and the number of IP addresses needed is 1,837 on average. Furthermore, after spending 24 seconds to check 3,000 addresses, the attacker has a probability of 80.0\% to identify a correct IP address that can be used to attack an arbitrary client. The measured probability is higher than the theoretical one, i.e., $1-(1-1/2048)^{3000}\approx76.9\%$.

To evaluate the threats of our off-path TCP attack on the Internet, we measure Alexa top 100k websites to identify how many websites suffer from the vulnerable IPID assignment. We observe that 22,953 websites are vulnerable to forged ICMP ``Fragmentation Needed'' messages from off-path attackers and thus can be tricked into clearing the \texttt{DF} flag of TCP packets and downgrading the IPID assignment. 
%
These websites are vulnerable to our attack\footnote{The attack may be disturbed by noises of non-TCP traffic that happens to share the same hash-based IPID counter with the attack traffic, however we measure that the actual disturbance is negligible, see Section~\ref{noises}}.
%
%
We cannot confirm the effectiveness of the attack against 22,803 websites that are unreachable from our vantage point in California due to packet filtering performed by the ISP hosting the vantage point. Moreover, we suspect that the rest resist to our attack due to two reasons, i.e., the OS versions of the websites are invulnerable (e.g., old Linux kernel versions or Windows), or the forged ICMP error messages are blocked.

\subsection{Results of TCP DoS Attacks}\label{implementation-reset}
In this experiment, we show that a TCP connection between a victim server and a victim client can be reset by an off-path attacker, resulting in a DoS attack. We conduct the attack under the common  scenario of SSH.

\noindent\textbf{Experimental Setup.} This attack involves 3 hosts, namely, an SSH server equipped with Ubuntu 18.04 (kernel version 4.18 or beyond), OpenSSH 7.6 and OpenSSL 1.0.2, a victim client who accesses the server based on SSH connections, and an attack machine equipped with Ubuntu 18.04 (kernel version 4.15) and a prefix of 152.136.0.0/16 that contains 2000 IP addresses in this prefix. The attack machine can use these IP addresses to detect hash collisions with the target client. The attacker attempts to reset the connection via sending TCP \texttt{RST} packets to the server. 

\noindent\textbf{Attack Procedure.} In this attack, the 3-tuple [client IP address, server IP address, server port] is known. First, the attacker identifies an IP address in its prefix which collides with the client IP address. Second, based on the identified attacker IP address, the attacker infers the correct client port number and the exact sequence number ($RCV.NXT$ on the server side) of the victim SSH connection. Finally, a spoofed \texttt{RST} packet specified with the inferred value is sent to the server, and the server will be tricked into resetting the SSH connection from the victim client. In this attack, the acceptable acknowledgment numbers are not needed.

\begin{table}[h]
\vspace{-2mm}
	\caption{Experimental results of SSH connection reset.}
	\vspace{-3mm}
	\begin{center}
		\begin{tabular}{@{}ll|ccc@{}} 
			\toprule
			\textbf{\begin{tabular}[c]{@{}c@{}}Server \\ address\end{tabular}} &
			\textbf{\begin{tabular}[c]{@{}c@{}}Linux\\ version\end{tabular}} & \textbf{\begin{tabular}[c]{@{}c@{}}Time \\ cost (s)\end{tabular}} & \textbf{\begin{tabular}[c]{@{}c@{}}Bandwidth \\ cost (KB/s)\end{tabular}} & \textbf{\begin{tabular}[c]{@{}c@{}}Success\\ rate\end{tabular}} \\ 
			\midrule
			62.234.203.$\times$ & 4.19 & 148.7 & 25.02 & 9/10  \\
		    152.136.49.$\times$  & 4.20 & 150.2 & 25.03 & 10/10 \\
			152.136.59.$\times$ & 5.3  & 160.1 & 24.95 & 9/10  \\
			62.234.182.$\times$ & 5.5  & 156.3 & 25.20 & 44/50 \\ 
			\bottomrule
		\end{tabular}
		\label{reset_results}
	\end{center}
\vspace{-3mm}
\end{table}

\noindent\textbf{Experimental Results.} On average, the time cost of identifying a correct attacker IP address is 15.4 seconds, and the correct one will be identified after checking 1,837 IP addresses on average. Table~\ref{reset_results} illustrates our experimental results. We test the attack against 4 servers that are equipped with Linux kernel version 4.19,  4.20, 5.3, and 5.5, respectively. The diversity of servers ensures the feasibility and effectiveness of the attack. The average time cost of resetting an SSH connection is 155 seconds, and the success rate is over 88\%. TCP connections DoS attack is particularly applicable to compromising applications secured by encrypted traffic, e.g., HTTPS and SSH.

\subsection{Results of TCP Manipulation Attacks}\label{inject-atc}
In this case, we perform two attacks to demonstrate that the newly discovered IPID side channel can be exploit to manipulate a TCP connection maliciously, thus causing serious damage to the upper applications including HTTP and BGP.

\noindent \textbf{(1) Manipulating Web Traffic.} We demonstrate that under the typical web application scenario, an off-path attacker can detect a victim client connecting to the target web server and then hijack the connection between the server and the client.

\noindent\textit{\bf Experimental Setup.} This attack involves 3 hosts. A web server is equipped with Linux kernel version 5.5 and a popular real-time communication web application called Rocket.Chat~\cite{Rocket}. An attack machine  is equipped with Ubuntu 18.04 (kernel version 4.15), and it is able to send packets to the server with a spoofed IP address. A client can access the web server based on HTTP. Note the OS type or version of the client is unrestricted in our attack. The attacker attempts to identify the potential victim client and hijack the TCP connection between the server and the client. For instance, the attacker may impersonate the victim client to inject malicious segments into the server and then inject fake messages into the chatting group. Here, the server IP address and server port are publicly known.

\noindent\textit{\bf Attack Procedure.} The attacker first downgrades the server's IPID assignment and detects potential victim clients who share the same hash-based IPID counter with the attacker. Next, the attack can be constructed  in the following four steps: (1) detecting whether the client has a TCP connection to the server, i.e., identifying correct source port number to obtain the TCP 4-tuple information, (2) inferring the exact sequence number, i.e., $RCV.NXT$ on the server, which can slide the server's receive window, leading to that the segment can be delivered to HTTP immediately, (3) inferring the acceptable acknowledgment numbers, and (4) injecting forged segments specified with the inferred values into the server and pushing fake messages into the chatting group.

\begin{figure}[h]
    \vspace{-4mm}
	\begin{center}
		\includegraphics[width=0.48\textwidth, height=2.1in]{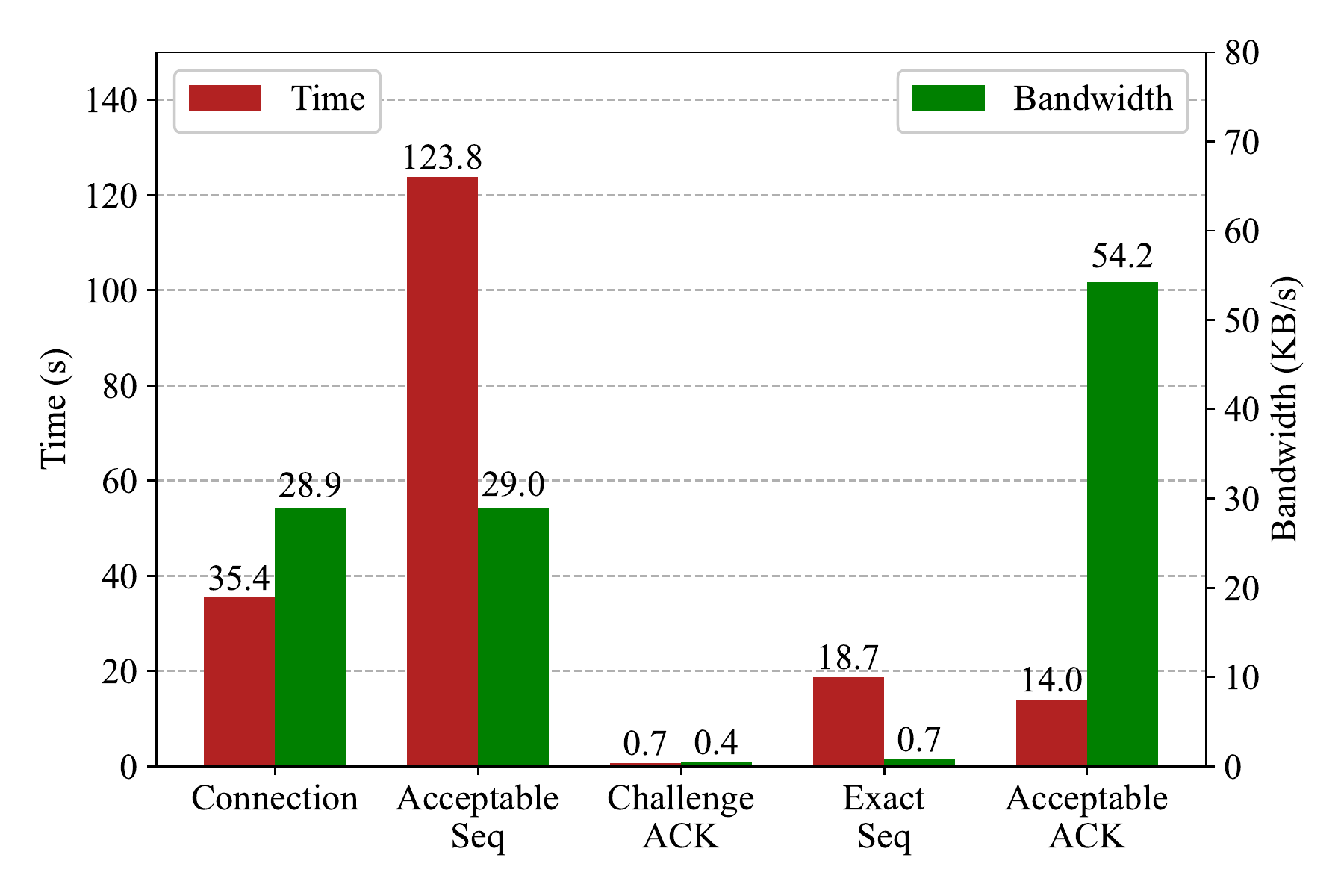}
		\vspace{-8mm}
		\caption{Time/Bandwidth overheads of web manipulation.}
		\label{time-overhead}
	\end{center}
	\vspace{-4mm}
\end{figure}

\begin{figure}[h]
    \vspace{-3mm}
	\begin{center}
		\includegraphics[width=0.35\textwidth, height=1.8in]{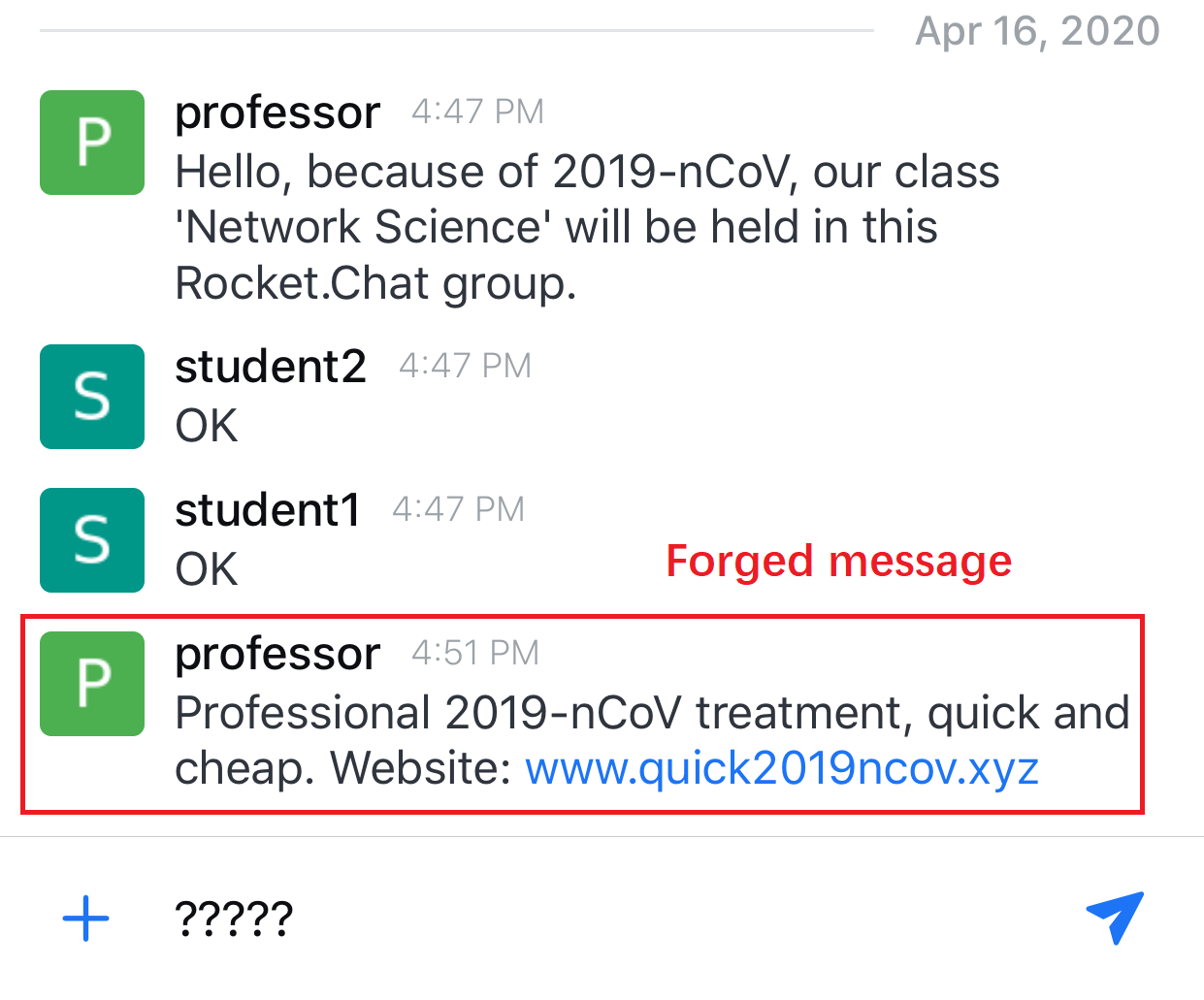}
		\vspace{-5mm}
		\caption{Snapshot of messages viewed by group members.}
		\label{web-poisoning}
	\end{center}
	\vspace{-4mm}
\end{figure}

\noindent\textit{\bf Experimental Results.} It takes 14.0 seconds to detect a potential victim client and 35.4 seconds to identify the correct source port number of the TCP connection. Figure~\ref{time-overhead} shows that the time cost in inferring the acceptable sequence numbers, locating the challenge \texttt{ACK} window, detecting the exact sequence number and an acceptable acknowledgment number are 123.8 seconds, 0.7 seconds, 18.7 seconds and 14.0 seconds, respectively. On average, the overall time cost of this attack is 206.6 seconds, including the time cost of detecting the victim client that can be performed in advance.  64.3\% of the overall time is spent on inferring the acceptable sequence numbers. The reason is that the server's receive window is relatively narrow and the attacker must sample a large number of sequence numbers. The average bandwidth overhead of this attack is 23.55 KB/s. Finally, when the server accepts the forged segment, the fake messages in the segment will be stored on the server and advertised to group members, as shown in Figure~\ref{web-poisoning}. Overall, the success rate of this attack is over 90\%.

\noindent \textbf{(2) Manipulating BGP Routing Table.} In this case, we demonstrate that an off-path attacker can manipulate BGP routing tables by performing our attacks on long-lived TCP connections. We show that the off-path attacker can pretend to be a legitimate BGP router and inject malicious BGP messages into other BGP peers, thus poisoning their BGP routing tables.

\noindent\textbf{Experimental Setup.} This attack involves 3 hosts.  A BGP server is equipped with Linux kernel version 5.5, listening on its port 179. A BGP client is equipped with Ubuntu 18.04. Both the server and the client run the BGP suite of Quagga~\cite{zebra} with version 1.2.0. After the client initiates a BGP connection, the two peers advertise BGP messages to each other and update their BGP route tables. An attack machine is equipped with Ubuntu 18.04,  and it is able to send packets with a spoofed IP address. The attacker aims to identify the potential victim client and hijack the BGP connection. We show that the attacker can impersonate the client and manipulate the server's BGP route table. We assume that the server IP address and server port are publicly known.

\noindent\textit{\bf Attack Procedure.} Similar to the HTTP hijacking attack, after downgrading the server's IPID assignment and identifying a victim client, the attacker first learn the presence of a BGP connection between the server and the identified client. Then, it infers the exact sequence number and an acceptable acknowledgment number to the server. Finally, the attacker sends forged BGP messages to the server based on the inferred values to poison the routing table. 

\begin{figure}[h]
	\vspace{-3mm}
	\begin{center}
		\includegraphics[width=0.46\textwidth]{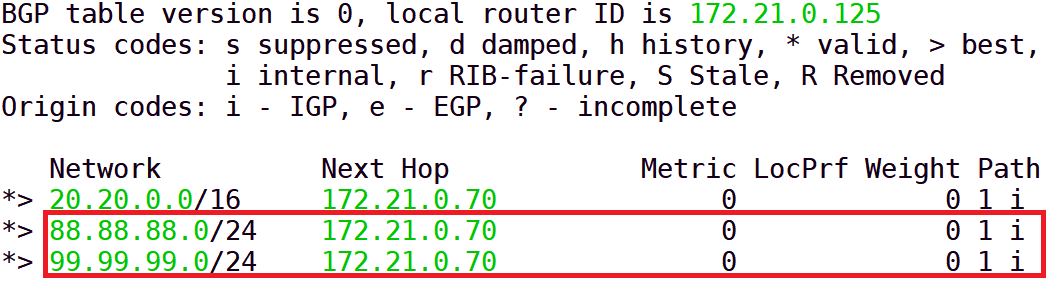}
		\vspace{-4mm}
		\caption{Snapshot of the poisoned BGP routing table.}
		\label{bgp-poisoning}
	\end{center}
	\vspace{-6mm}
\end{figure}

\noindent\textit{\bf Experimental Results.} Figure~\ref{bgp-poisoning} presents a snapshot of the poisoned BGP routing table. The Network Layer Reachability Information (NLRI) of network ``99.99.99.0/24'' and ``88.88.88.0/24'' is fake, which are not advertised by router 172.21.0.70 but injected by the attacker. On average, the attacker can finish BGP routing table poisoning in 214.3 seconds (including the time cost of identifying the victim client), with a success rate over 90\%.

\section{Discussion and Countermeasure}
\label{sec:Discussion}

\subsection{Attack Robustness}\label{noises}


\noindent \textbf{(1) IPID noises.} Unlike the global IPID counter that is noisy due to the sharing between all outgoing traffic, the downgraded hash-based IPID counter shared between the attacker and the victim client on the TCP server side is reliable, and hence our attack does not suffer the traditional noise issue~\cite{ensafi2010idle,ensafi2014detecting,pearce2017augur,pearce2018toward}.

Since the outgoing TCP traffic directed to irrelevant clients uses per-socket-based IPID counters instead of the hash-based IPID counters, irrelevant TCP traffic will not disturb our attack.
%
%
We measure and evaluate the disturbances from the noises of non-TCP traffic in the real world and find that the impact is also limited. We find that the hash-based IPID counters of more than 91\% vulnerable websites in the Alexa top 100k websites list are not disturbed at all (within 5 minutes in our experiment), which means the IPIDs are always contiguous and there is no outgoing traffic sharing the same hash-based IPID counter with the attacker at the server in this time window. 
Note that, less than 9\% vulnerable websites are disturbed by non-TCP traffic, e.g., the ICMP traffic generated by these websites, that happens to share the same hash-based IPID counter with the attack traffic during the period.
%
Considering that our attack can be finished within 215 seconds, the real disturbance is negligible.
Moreover, other types of noises that specific to certain application/network scenarios, e.g., packet loss, can be effectively mitigated by re-running the attack multiple times. For example, in our experiments, when we detect a potential victim, we usually conduct the detection process again to enhance the confidence, which incurs around 6 seconds additional delay but can almost eliminate the false positives. We confirm the results by generating random packet loss in our experiments.

\noindent \textbf{(2) Multiple Connections from the Victim Client.} The potential victim clients who collide with the attacker can be identified in advance (i.e., being independent of TCP connections), so the side channel will always exist until the server restart.

When there is more than one TCP connection between the server and the victim client, since these connections are mapped into the same hash-based IPID counter, they will interfere with the attacker's observation, thus affecting the success rate of the attack against the target connection. However, in practice, this circumstance is rare, since the client usually does not initiate multiple TCP connections to the same server in parallel, particularly,  HTTP/1.1~\cite{rfc2616} and HTTP/2~\cite{rfc7540} work based on one TCP connection even if there are multiple HTTP requests to the server.  

\noindent \textbf{(3) More Victim Clients under the NAT Scenario.} Network Address Translation (NAT) is a widely used technique to overcome the shortage of IPv4 addresses~\cite{rfc2663}. Under this  scenario, multiple hosts share a public IP address. As a result, if the attacker identifies a potential victim client who accesses the Internet via NAT technique, it indicates that all hosts behind the same NAT gateway are potential victims. Hence, in practice, the actual number of victim clients is far greater than the number of being identified, and the NAT technique incurs a more wide attack surface. 

Note if more than one host behind the NAT gateway connects to the target server at the same time, the TCP connections initiated from these hosts will share the same IPID counter on the server and would interfere with the attacker's observation. In practise, the impact is limited, since it requires these hosts to access the same target server at the same time and all the established connections have continuous data transmission from the server to the hosts.

\noindent \textbf{(4) Shifting Sequence and Acknowledgment Numbers.} Another circumstance that may affect the success rate of the attack is the shifting of the sequence and acknowledgment numbers, i.e., if the victim TCP connection has ongoing traffic, the acceptable sequence and acknowledgment numbers may shift as the attack is in progress. This problem can be solved by the repeated inference of the acceptable sequence and acknowledgment numbers. We confirm that if the receive window does not slide very quickly, e.g., under the scenarios of SSH and BGP, the success rate of the attack will not be affected obviously. Even if the receive window slides quickly enough to break the attacker's inference, the attacker can choose to exploit the other side of the TCP connection where the receive window slides more slowly.

\vspace{-0.1in}
\subsection{Countermeasures}
We have reported the newly discovered IPID side channel to the Linux community. Meanwhile, we also propose to throttle the exploit via eliminating the root cause, i.e., repairing the IPID assignment to TCP packets.

\noindent \textbf{(1) Assigning IPID Based on the \texttt{Protocol} Field.} The root cause of the attack is that Linux can be tricked into choosing an incorrect IPID assignment policy for TCP packets. When Linux assigns IPID to TCP packets, it decides which policy to be chosen based on the \texttt{DF} flag in IP header, rather than the \texttt{Protocol} field. Therefore, attackers can clear the \texttt{DF} bit of the TCP packets by forging ICMP ``Fragmentation Needed'' messages, which causes hash collisions and build a side channel. To address this issue, we propose to assign IPID by evaluating if a packet is originated from TCP based on the field of \texttt{Protocol} in IP header, instead of the \texttt{DF} flag. If the packet's \texttt{Protocol} field is specified as TCP, we assign IPID for the packet based on the per-socket assignment policy. As a result, all TCP packets will no longer share IPID counters with the attacker and the side channel can be eliminated. We implement the mechanism in Linux 4.18 and confirm its effectiveness through real evaluation.

\noindent \textbf{(2) Enhancing IPID Assignment for \texttt{RST} Packets.} Another countermeasure is to change the IPID assignment of \texttt{RST} packets, which allows an attacker to infer the connection information. Since Linux kernel version 4.18, Linux directly sets the IPID of \texttt{RST} packets to 0, which is also vulnerable and can be abused. When an attacker learn the presence of a TCP connection between the server and the identified victim client, the attacker can forge \texttt{SYN/ACK} packets. If there is no connection initiated from the specified source port, the server responds with a \texttt{RST} packet, otherwise, with a challenge \texttt{ACK} packet. The IPID of the \texttt{RST} packet is 0, and hence it will not cause an increment to the shared IPID counter. It will be different from the behavior of the challenge \texttt{ACK} packet, which enables an indicator for the attacker to judge the existence of the connection. 

Thus, we propose to modify the IPID assignment for \texttt{RST} packets. Note we cannot assign IPID for \texttt{RST} packets based on a socket preserved counter, since the \texttt{RST} packets may be generated and issued independently of a TCP connection. Also, we cannot assign IPID for \texttt{RST} packets based on hash IPID counters, since this assignment is vulnerable to TCP/IP connections detecting attacks~\cite{alexander2019detecting}. An empirical method is to assign IPID for \texttt{RST} packets based on the destination of the packet, i.e., selecting the IPID counter assigned to the destination. If there is a TCP connection to the destination (the victim client in our scenario) and the counter preserved in the socket will be selected, it can avoid the differences on the counter. 


\section{Related Work}
\label{sec:related}

\noindent \textbf{IPID Side Channels.} 
IPID ensures the uniqueness of a packet for packet fragmentation and reassembly~\cite{RFC791,rfc6864}. However, IPID has been widely abused to conduct off-path attacks due to the vulnerable assignment methods. Ensafi \etal performed idle port scan and network protocol analysis by leveraging the side channel of global IPID counters~\cite{ensafi2010idle}. They also suggested that the global IPID counters can be used to detect intentional packet drops~\cite{ensafi2014detecting}.
By leveraging the side channel of global IPID counters, Pearce \etal measured the reachability between any two Internet locations without controlling a measurement vantage point~\cite{pearce2017augur,pearce2018toward}. 
Jeffrey \etal showed that per-destination IPID counters are also vulnerable, which can be exploited to infer the number of packets between two machines with UDP and ICMP and even learn the presence of a TCP connection by launching off-path attacks~\cite{Knockelcounting}. 
Alexander \etal detected TCP connections via IPID hash collisions. They leveraged the IPID of the triggered \texttt{RST} packets to determine the presence of a victim TCP connection~\cite{alexander2019detecting}. Their method can only detect TCP connections, but not hijacking a TCP connection. Moreover, the vulnerabilities they used have been fixed since Linux kernel version 4.18. In this paper, we identified a new vulnerability of abusing IPID, which can be exploited by off-path attackers to perform a TCP hijacking attack.

\noindent \textbf{TCP Hijacking Attacks.}
Cao \etal found that an off-path attacker can infer whether two arbitrary hosts on the Internet are communicating using a TCP connection by utilizing a side channel in the challenge \texttt{ACK} mechanism, identify the sequence and acknowledgment numbers of the connection, and then hijack the connections~\cite{cao2018off,cao2016off}. The side channel vulnerability has been eliminated by setting a random challenge \texttt{ACK} count limit. 
A timing side channel has been uncovered in the half-duplex IEEE 802.11 or Wi-Fi technology, which can be exploited by an off-path attacker to inject data into a TCP connection and force the browser to cache malicious objects~\cite{chen2018off}. 
By exploiting the global IPID counter which was adopted by the previous Linux and Windows systems, Gilad \etal inferred if two hosts have established a TCP connection identified by a specific four-tuple and then launch off-path TCP injection attacks to poison the HTTP and Tor traffic~\cite{gilad2014off,gilad2012spying,gilad2013off,gilad2012off}.

Besides, unprivileged applications or sandboxed scripts controlled by attackers running on victim hosts (called puppets) can also be  leveraged to perform off-path TCP attacks~\cite{qian2012collaborative, qian2012off,gilad2013tolerance}. 
Qian \etal uncovered that the middlebox of firewall can be abused to perform the TCP sequence number inference attack~\cite{qian2012off}, and conducted a collaborative TCP sequence number inference attack by exploiting the packet counter side channels~\cite{qian2012collaborative}.
Gilad \etal identified that attackers can conduct web cache poisoning attacks by leveraging a restricted script in the user's browser sandbox~\cite{gilad2013tolerance}. 
Compared with these attacks, our off-path TCP attack is a pure off-path one and does not need any assistance of puppets. Moreover, our attack leverages a new side channel vulnerability appearing in the interactions among IP, ICMP, and TCP, which cannot be unearthed by the principled methods~\cite{cao2019principled}.

\noindent\textbf{TCP DoS Attacks.}
TCP \texttt{SYN} flooding is a major threat that is difficult to be identified due to the similarity to the legitimate establishment of TCP connections~\cite{rfc4987,bani2013syn,wang2002detecting,aborujilah2014detecting}. Similarly, TCP \texttt{FIN} flooding, TCP \texttt{RST} flooding, TCP \texttt{ACK} flooding, TCP \texttt{URG} flooding, and TCP \texttt{Null} flooding are other typical TCP DoS attacks~\cite{acharya2016survey,35DDoS}, which aim to cause resource exhaustion of the TCP connection. Besides these brute-force attacks, more sophisticated and stealth DoS attacks against TCP protocol have been developed and become difficult to detect or mitigate, such as low-rate TCP-targeted DoS attacks~\cite{kuzmanovic2003low, shevtekar2005low, herzberg2010stealth,jero2018automated} or congesting intermediate links~\cite{smith2018routing,MuoiTran2019routing}. 


\section{Conclusion}
\label{sec:conclusion}
In this paper, we uncover a new off-path TCP hijacking attack that leverages a subtle side channel in the new mixed IPID assignment method of Linux kernel version 4.18 and beyond. We find that a pure off-path attacker can downgrade the IPID assignment for TCP packets from the more secure per-socket-based policy to hash-based policy, thus building a shared IPID counter that can be observed to learn the presence of victim TCP connections and then infer the sequence numbers and acknowledgment numbers of the connection. We evaluate the impacts of the new off-path TCP exploit on the Internet and implement the exploit under different scenarios. Our experiments show that off-path attackers can perform various attacks by exploiting the newly discovered IPID side channel, e.g., resetting SSH connections, manipulating web traffic and poisoning BGP routing tables. We also propose to eliminate the root cause of the exploit via repairing the IPID assignment to TCP packets in Linux. We implement our countermeasure and confirm its effectiveness in practice.


\begin{acks}
We thank the anonymous reviewers for their insightful comments. We are grateful to our shepherd Paul Pearce for his guidance on improving our work.
This work is supported by the National Key R\&D Program of China with No.2018YFB1800402, National Science Foundation for Distinguished Young Scholars of China with No.61825204, National Natural Science Foundation of China with No.61932016, U1736209, and 61572278, Beijing Outstanding Young Scientist Program with No.BJJWZYJH01201910003011, U.S. ONR grants N00014-16-1-3214 and N00014-18-2893, U.S. ARO grant W911NF-17-1-0447, and the project "PCL Future Greater-Bay Area Network Facilities for Largescale Experiments and Applications (LZC0019)".
Ke Xu is the corresponding author of this paper.
\end{acks}

\bibliographystyle{ACM-Reference-Format}
\bibliography{reference}



\end{document}